\documentclass[iop]{emulateapj}

\newcommand{\sref}[1]{Section \ref{#1}}
\newcommand{\fref}[1]{Fig. \ref{#1}}
\newcommand{\tref}[1]{Table \ref{#1}}

\usepackage{epsfig}
\usepackage{subfigure}

\shorttitle{Galaxy satellites in SDSS}
\shortauthors{Lares et al.}

\begin{document}

\title{Properties of satellite galaxies in the SDSS photometric survey:\\
luminosities, colours and projected number density profiles.}

\author{Marcelo~Lares, Diego~Garc\'{\i}a~Lambas and Mariano~Dom\'{\i}nguez}
\affil{Instituto de Astronom\'{\i}a Te\'{o}rica y Experimental
  (CONICET-UNC). Observatorio Astron\'{o}mico de C\'{o}rdoba (UNC), 
  Laprida 854, X5000BGR, C\'{o}rdoba, Argentina}  
\email{mlares@mail.oac.uncor.edu}
%
%\author{Diego~Garc\'{\i}a~Lambas}
% \affil{Instituto de Astronom\'{\i}a Te\'{o}rica y Experimental
%  (CONICET-UNC). Observatorio Astron\'{o}mico de C\'{o}rdoba (UNC), 
%  Laprida 854, X5000BGR, C\'{o}rdoba, Argentina}  
%
%\author{Mariano~Dom\'{\i}nguez}
%\affil{Instituto de Astronom\'{\i}a Te\'{o}rica y Experimental
%  (CONICET-UNC). Observatorio Astron\'{o}mico de C\'{o}rdoba (UNC), 
%  Laprida 854, X5000BGR, C\'{o}rdoba, Argentina} 

%%%%%%%%%%%%%%%%%%%%%%%%%%%%%%%%%%%%%%%%%%%%%%%%%%%%%%%%%%%·B·E·G·I·N·

\begin{abstract}
    We analyze photometric data in SDSS-DR7 to infer statistical
    properties of faint satellites associated to isolated bright
    galaxies  ($M_r<-20.5$) in the redshift range $0.03<z<0.1$.
    The mean projected radial number density profile shows an excess
    of companions in the photometric sample around the primaries, with
    approximately a power law shape that extends up to $\simeq 700$ kpc.
    Given this overdensity signal, a suitable background subtraction
    method is used to study the statistical properties of the
    population of bound satellites, down to magnitude $M_r=-14.5$, 
    in the projected radial distance range 
    \mbox{$100<r_p/\rm{kpc}<3<R_{vir}>$}.
    The maximum projected distance corresponds is in the
    range $470-660$ kpc for the different samples.
    We have also considered a colour cut consistent with the observed
    colours of spectroscopic satellites in nearby galaxies so that
    distant redshifted galaxies do not dominate the statistics.
    We have tested the implementation of this background subtraction
    procedure using a mock catalogue derived from the Millenium
    simulation SAM galaxy catalogue based on a $\Lambda$CDM model.
    We find that the method is effective in reproducing the true
    projected radial satellite number density profile and luminosity
    distributions, providing confidence in the results
    derived from SDSS data. 
    We find that the spatial extent of satellite systems is larger for
    bright, red primaries. Also, we find a larger spatial distribution
    of blue satellites.  
    For the different samples analyzed, we derive the average number
    of satellites and their luminosity distributions down to
    \mbox{$M_r=-14.5$}.
    The mean number of satellites depends very strongly on host
    luminosity. Bright primaries \mbox{($M_r<-21.5$)} host on average
    $\sim 6$ satellites with \mbox{$M_r<-14.5$}.
    This number is reduced for primaries with lower luminosities
    \mbox{($-21.5<M_r<-20.5$)} which have less than $1$ satellites per
    host.
    We provide Schechter function fits to the luminosity distributions
    of satellite galaxies where the resulting faint end slopes equal 
    to  $−1.3 \pm 0.2$, consistent with the universal value.
    This shows that satellites of bright primaries lack an excess
    population of faint objects, in agreement with the results in
    the Milky Way and nearby galaxies.
\end{abstract}

%··········································································
%                                00SECTIONS

\section{Introduction} \label{S_introduction}
%{{{/*

According to the currently accepted model of structure formation, galaxy
systems arise as the result of hierarchical clustering
\citep{white_galaxy_1991, bertschinger_cosmic_1994, cole_recipe_1994}.
The details by which galaxies form and evolve in dense or moderately dense
environments, where galaxy-galaxy interactions are frequent and matter
distributes in a rich substructure, depend on the characteristics of those
environments.
The assembly of galaxy systems entail the process of matter accretion,
governed by gravity, as well as astrophysical phenomena, such as the
efficiency of gas to cool and collapse or the energy feedback related to
the late stages of stellar evolution
\citep{viola_how_2008,kang_massive_2006}.
Some of these details are not yet fully understood, and observational
evidence is fundamental to constraint structure formation and evolution
models, specially because faint galaxies are more sensitive to
astrophysical processes like supernova feedback and ram pressure stripping.
In particular, statistical studies of systems of galaxies are key to
understand the transformations of galaxies due to the interactions between
galaxies and their environment \citep{nichol_interplay_2003}.

Although the formation of a galaxy is believed to take place on the
potential well of a dark matter halo, not all haloes host a galaxy.
As pointed out first by \citet{klypin_where_1999}, the number of halos
in numerical simulations were an order of magnitude greater than the
observed number of satellites in the local group
\citep{moore_dark_1999, kravtsov_tumultuous_2004,
strigari_redefining_2007}.
This difference has been a matter of lively debate, being attributed
either to a lack of observed galaxies or to an excess of formed
objects in simulations.
\citet{willman_sdss_2002} discussed the possibility of under counting
satellite galaxies in the Milky Way and estimated the actual number of
satellites in about twice the known population at that time.
The authors argued that galactic extinction and stellar foreground can
lead up to 33 per cent of incompleteness, and then the number of Milky Way
satellites at low galactic latitude and at galacto--centric distance
might be underestimated.
\citet{simon_kinematics_2007} found that the number of satellites in
the Milky Way system was greater than previously expected, based on an
analysis of the SDSS data.
With these findings, the discrepancy between the number of observed
and simulated satellites reduces to a factor of nearly 4.
Nevertheless,
applying a background subtraction technique
on photometric data from SDSS-DR7,
\citet{liu_how_2010} found recently 
that the Milky Way galaxy has significantly more
bright satellites than a typical galaxy of its luminosity.

The study of the spatial distribution of satellites around primaries
and clusters has become favored by increasingly large galaxy redshift
surveys.
Many works address observational studies of radial distribution of
galaxies in spectroscopic samples \citep{coil_deep2_2006,
lin_kband_2004, yang_cross_2005, collister_distribution_2005}, mostly
around galaxy clusters, and on bright primary galaxies
\citep{sales_satellite_2005,chen_constraining_2006}.
Deeper samples have been also used to compute projected density
profiles, using background subtraction \citep{hansen_measurement_2005}
around MaxBCG galaxy systems.
Moreover, galaxy projected density profiles were computed based on
projected correlation function determinations in redshift galaxy
catalogues \citep{li_luminosity_2007} and deeper photometric samples
\citep{wang_galaxy_2010} around a set of spectroscopically identified
galaxies.

The distribution of satellite luminosities is also key to the development
of models and understanding of the processes of galaxy formation
\citep{benson_what_2003,benson_galaxy_2010,okamoto_properties_2010}.
Given the low luminosity of most satellites, however, their observation is
usually onerous and hardly accessible outward of the Local Group.
\citet{mateo_dwarf_1998} put forward a detailed census of dwarf galaxies,
from which a flat faint end of the luminosity function in the Local Group
could not be discarded.
Background subtraction methods have been widely used to obtain galaxy
luminosity function in clusters,
and also results of this procedure on individual clusters have been
reported
\citep[e.g.][]{oemler_systematic_1974}.
Since it is not limited to the computation of luminosities, it can be also
used to obtain the color-magnitude relation \citep{pimbblet_are_2008}.
\citet{andreon_rigorous_2005} present a variation of the background
decontamination method, avoiding the use of arbitrary binning and
incorporating the background noise as part of a refined model for the
description of data.
\citet{koposov_luminosity_2008} presented a search methodology for Milky
Way satellite galaxies in SDSS data through the computation of efficiency
maps.
Search for stellar concentrations using these maps suggest a luminosity
distribution that steadily rises following a power law up to $M_v\simeq 5$.
From there on, a flat distribution could not be discarded
\citep{koposov_luminosity_2008}.

\citet{tollerund_hundreds_2008} use completeness limits for the SDSS-DR5 to
implement a correction for luminosity bias.
Although a first order correction would produce an increase in the faint
end of the luminosity function, the authors bring forward that this result
is not well enough constrained given available data.
\citet{trentham_faint_2002} study the faint end of the galaxy luminosity
function in five different local environments, from the Virgo cluster to
NGC 1023 group.  The authors derive an averaged luminosity distribution in
the range $-18.<M_r<-10$ (Cousins R magnitude) and infer a faint end slope
$\alpha\sim -1.2$.
In the particular case of NGC 1023, a more detailed study confirmed later
that the faint end is consistent with a shallow slope
\citep{trentham_dwraf_2009}.
\citet{tully_midlife_2008} studied the NGC 5353 group and attributed a
faint end slope $\alpha=-1.15$ to the fact that this group is at an
intermediate evolutionary age.

Membership of individual galaxies through spectroscopic measurements is
inefficient in terms of observing time, given the large fraction of
background objects that have to be rejected. 
This results in few systems with derived luminosity function complete down
to faint magnitudes.
For this reason, a background subtraction technique is an efficient method
to study, on a statistical basis, properties of the population of companion
objects.
Using deep mock catalogues constructed from a numerical simulation,
\citet{valotto_clusters_2001} analyze systematic effects in the
determination of the galaxy luminosity function in clusters.
Their results indicate a strong tendency to derive a rising faint end when
clusters are selected without redshift information.
This is due to projection effects, since many of the clusters selected in
2D have no significant counterpart in 3D.
\citet{munoz_accuracy_2009} use Mock catalogues constructed using GALFORM
\citep{baugh_primer_2006} semianalytic model of galaxy formation to study
the reliability of the statistical background subtraction method to recover
the underlying observer-frame luminosity function of high redshift
($z\simeq1$) cluster galaxies in the K$_s$ band.
These authors find that the optimal response of the method in recovering
the underlying galaxy luminosity function occurs when background corrected
counts of faint galaxies are complemented with photometric redshifts of
bright galaxies.
They also show that the increase in the number of galaxy clusters that
contribute to the computations dramatically reduce stochastic errors.
\citet{christlein_dependence_2000} study luminosity functions for galaxies
in loose groups and suggests that the ratio of dwarf to giant galaxies is
continuously increasing from low to high mass groups.
This is based on data from the Las Campanas Redshift Survey, where
environment is estimated using the line-of-sight velocity dispersion of the
host groups.
Background subtraction has been applied to single clusters
\citep{andreon_rigorous_2005, barkhouse_luminosity_2007} and to ensembles
of clusters \citep{gonzalez_faint-end_2006}.

%--------------------------------------------------------- FIGURE 1 -
\begin{figure} \centering
   \includegraphics[width=0.49\textwidth]{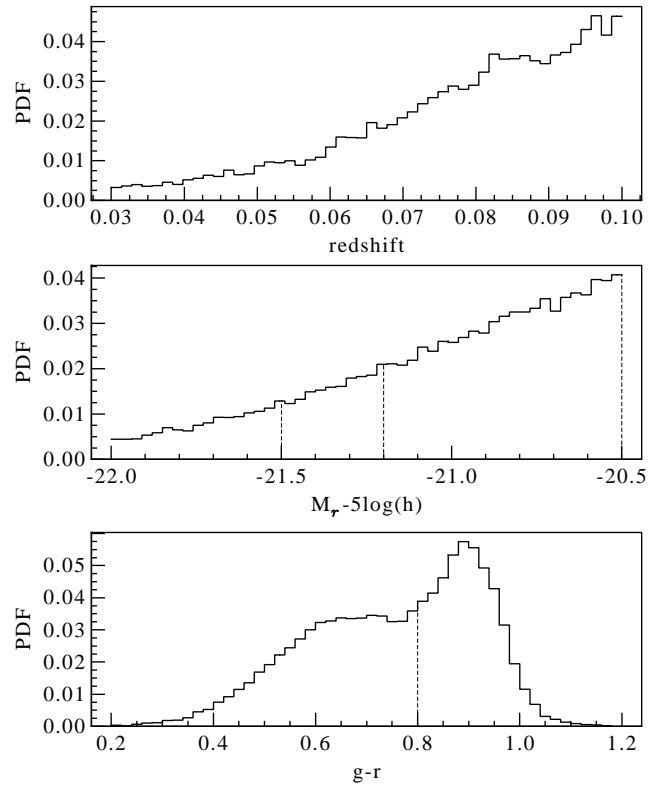}
   \caption{Properties of primary galaxies in the total sample.  This
   sample comprises $51710$ galaxies brighter than $M_r=-20.5$ in the
   redshift range $0.03$ to $0.1$, satisfying the isolation criteria
   stated in \sref{S_data:hosts}. } \label{F_PrimaryProps}
\end{figure}% 
%-------------------------------------------------------------------- 

The goal of this work is to obtain statistical properties of satellite
galaxies in the magnitude range $-18.5<M_r<-14.5$, using SDSS
photometric data.
Previous studies of galaxy satellites concern mostly Milky Way and
nearby galaxies, so that this work complements with a study of
projected radial number density profiles, luminosity, and colour
distributions for a statistically large sample of primaries within
$z=0.1$.
The definitions of host samples and the adopted satellite selection
criteria are presented in \sref{S_data}.
The details of the background subtraction procedure are given in
\sref{S_method}.
Then, in \sref{S_results} we present the derived distributions of
satellite properties.
In order to validate the implemented method, we test it in
\sref{S_testing}.
Finally, the main conclusions are provided in \sref{S_conclusions}.
%
%}}}/*

\section{Photometric and spectroscopic data} \label{S_data}
%{{{/*

%------------------------------------------------------------ TABLE 1
\begin{table*} \begin{minipage}{\textwidth} \centering

\begin{tabular*}{\textwidth}{@{\extracolsep{\fill}}cccccc}
\hline
 & \multicolumn{3}{c}{primaries}& \multicolumn{2}{c}{galaxies} \\
\cline{2-4} \cline{5-6} sample &luminosity &colour & $N_{p}$
&$R_{max}$ [kpc]  &colour\\ 
\hline

S0-A-a &  M$_r<$-20.5   &     all     & 51710 & $<$ 480 & $-0.4<g-r<1.0$  \\ 
S0-A-r &  M$_r<$-20.5   &  $g-r>1.0$  & 51710 & $<$ 480 &$0.4<g-r<1.0$    \\ 
S0-A-b &  M$_r<$-20.5   &  $g-r<1.0$  & 51710 & $<$ 480 & $-0.4<g-r<0.4$  \\ 
\\ 
S1-A-a &-21.5$<M_r<$-20.5&    all & 42335 & $<$     470 & $-0.4<g-r<1.0$  \\ 
S1-A-r &-21.5$<M_r<$-20.5& $g-r>1.0$  & 42335 & $<$ 470 &$0.4<g-r<1.0$    \\ 
S1-A-b &-21.5$<M_r<$-20.5& $g-r<1.0$  & 42335 & $<$ 470 & $-0.4<g-r<0.4$  \\
\\ 
S2-A-a &  M$_r<$ -21.5  &     all     & 9375  & $<$ 660 & $-0.4<g-r<1.0$  \\ 
S2-A-r &  M$_r<$ -21.5  &     all     & 9375  & $<$ 660 &$0.4<g-r<1.0$    \\ 
S2-A-b &  M$_r<$ -21.5  &     all     & 9375 & $<$  660 & $-0.4<g-r<0.4$  \\ 
\\ 
S2-R-a &  M$_r<$ -21.5  & $g-r>1.0$  & 4740  & $<$  660 & $-0.4<g-r<1.0$  \\ 
S2-R-r &  M$_r<$ -21.5  &  $g-r>1.0$  & 4740  & $<$ 660 &$0.4<g-r<1.0$    \\ 
S2-R-b & M$_r<$ -21.5  &  $g-r>1.0$  & 4740 & $<$   660 & $-0.4<g-r<0.4$  \\ 
\\
S2-B-a &  M$_r<$ -21.5  &  $g-r<1.0$  & 4635  & $<$ 660 & $-0.4<g-r<1.0$  \\ 
S2-B-r &  M$_r<$ -21.5  &  $g-r<1.0$  & 4635  & $<$ 660 &$0.4<g-r<1.0$    \\ 
S2-B-b &  M$_r<$ -21.5  &  $g-r<1.0$  & 4635 & $<$  660 & $-0.4<g-r<0.4$  \\
\hline
\end{tabular*} 

\caption{%
   Definition of satellite samples considering colour range, maximum
   projected distance to the primary and primary properties.
   The number of primaries $N_p$ of each sample is also indicated.
   Satellite galaxies have a lower bound of $g-r=-0.4$ for samples
   {\it a} and {\it b}.
   Upper case letters A, B and R stand for All, Blue and Red,
   respectively.  Similarly, lower case letters indicate the ranges of
   satellite colours.} 
\label{T_SampleDefs} \end{minipage}
\end{table*} 
%-------------------------------------------------------------------- 

We use the large database provided by the Sloan collaboration
\citep[SDSS,][]{stoughton_sloan_2002}, which provides photometric
information of objects down to faint magnitudes.
This survey has been carried out using a dedicated $2.5$m telescope
\citep{gunn_2.5_2006}, and comprises digital photometric information of
stars and galaxies in $5$ bands \citep{fukugita_sloan_1996,
smith_ugriz_2002} reduced by an automated pipeline
\citep{lupton_SDSS_2001}.
The limiting magnitude is $22.2$ in the \textit{r}--band.
A set of the brightest and more concentrated galaxies in the main galaxy
sample has been selected for spectroscopic follow up
\citep{blanton_efficient_2003}.
This leads to the spectroscopic galaxy catalogue, which contains galaxies
with Petrosian magnitude $r<17.77$ \citep{strauss_spectroscopic_2002}.
Both photometric and spectroscopic data are accessible through a web
interface to the public releases of the survey.
The Data Release 7 comprises nearly 18 Tb of catalogued data for objects
identified as galaxies by the automated data reduction pipeline \citep[Data
Release 7,][]{abazajian_seventh_2009}.
The main galaxy sample comprises $691055$ galaxies with 95 per cent
completeness down to a limiting magnitude \mbox{$r_{lim}=17.77$} in the
\mbox{$r$-band}.
The surface brightness limits are imposed by the instrument capabilities
and the automated reduction pipeline, so that these data sets allow to
retrieve information about galaxies above the surface brightness limit of
the catalogue, $\mu_{50} \leq 24.5\; mag\; arcsec^{-2}$ in the Petrosian r-band
\citep{strauss_spectroscopic_2002}.
All galaxies in this sample have redshift measurements and serve
in this work as primary
targets for the study of fainter galaxies, accessed from a deeper
photometric sample.
We obtain statistical properties of faint satellites, most of
them not present in the spectroscopic survey, associated to primaries with
measured redshifts in the range 0.03 to 0.1
To this aim, we use the New York Value Added Galaxy Catalog
\citep[NYU-VAGC,][]{blanton_new_2005} to extract photometric information of
neighboring galaxies of primaries.
This catalogue is based on the sixth release of the Sloan Digital Sky
Survey, and covers \mbox{$9583$ deg$^2$} of sky distributed into a northern
cap and three southern stripes.
We notice that the sky coverage of the main galaxy sample in DR7 is
approximately included in the NYU-VAGC DR6 area, so that the
cross-correlation between these two catalogues is suitable for the purpose
of this work.
We used software HEALPix \citep{gorski_healpix_2005} to optimize the
data processing of the galaxy catalogues.

%}}}/*

\subsection{Host samples and galaxy selection criteria} \label{S_data:hosts}
%{{{/*

We have considered primaries brighter than \mbox{$M_p=-20.5$}
\mbox{($r$--band} luminosities) in the redshift range $0.03$ to $0.1$,
applying an isolation criterion in order to avoid high density environments
such as pairs or groups of galaxies.
The density contrast around galaxies fainter than \mbox{$M_p=-20.5$} is low
and comparable to Poisson uncertainty.
Since the method, detailed in \sref{S_method}, is based on the presence of
a strong signal to background, we left them out of consideration in
defining the samples of hosts.
These primaries are constrained to have no neighbors brighter than $M_p+2$
within projected distance \mbox{$700$ kpc} and relative radial velocity
difference \mbox{$700$ km/s.}
These criteria are similar to those adopted in previous studies using
spectroscopic samples \citep{sales_satellite_2005, chen_constraining_2006,
agustsson_anisotropic_2010}, and are intended to select and isolate halos
where a dominating, assumed central galaxy of the satellite system is
found.
The total sample of primaries comprises $51710$ objects brighter than
\mbox{$M_p=-20.5$} in the redshift range $0.03$ to $0.1$.

Taking into account galaxy luminosities and colours, we have considered
different subsamples of primaries in order to explore possible dependencies
of the satellite properties as a function of host properties.
The description of the subsamples considered is given in
\tref{T_SampleDefs}.
Subsample names are indexed on a three character basis: a number indicating
the host luminosity selection (''0'' for full sample, ''1'' for hosts with
$-21.5<M_r<-20.5$, and ''2'' for hosts with $M_r<-21.5$); and two letters
indicating host and satellite colours.
For simplicity, uppercase characters correspond to primary galaxy colours
(A,R,B: All, Red, Blue), while lowercase indicate satellite colours (a, r,
b standing for all, red and blue respectively).
In \fref{F_PrimaryProps} we show the distributions of \mbox{$r$--band}
absolute magnitudes, \mbox{$g-r$ colours} and redshifts of hosts in the
total sample.
The bimodal distribution of primary galaxy colours can be clearly
appreciated, and we use a colour cut \mbox{$g-r$=0.8} to separate
subsamples according to host colour.

%}}}/*

\section{Background subtraction method} \label{S_method}
%{{{/*

The background subtraction methodology are based on the simple idea of
counting the number of objects in a region where a given signal is
expected to lie, superposed to an uncorrelated noise, and subtracting
a statistical estimation of that noise.
In this case, the signal is due to the presence of satellite galaxies
in systems dominated by a central and luminous galaxy, and the noise
is associated to the background and foreground galaxies not
dynamically linked to the primary galaxy.
Then, it allows to statistically obtain properties of the faint
galaxies associated to the primaries, without the need of redshift
information for individual objects.
This is accomplished provided that the working hypothesis of the
central primary is satisfied, and convenient ranges of observed
parameters are chosen, so that to minimize the contribution of
background counts.
Although the method does not allow to quantify the contribution to the
signal of each individual object, it is possible to obtain statistical
estimates of probability density distributions describing galaxy
properties.
This is accomplished by restricting galaxies to a fixed bin of the
variable, for instance the luminosity of the satellite or their
distance to the central galaxy, and normalizing to the number of
contributing systems in that bin, after implementing the background
subtraction.

Within the hierarchical clustering paradigm, the matter distribution can be
roughly described by a set of halos populated with galaxies, according to
certain recipes that depend on galaxy type \citep{cooray_halo_2002}.
This model can give insight to adopt an appropriate choice of the centers,
which is key to increase the signal from the satellite population against
the background noise and obtain reliable results using this method.
Central galaxies play a different role than the rest of objects within the halo due
to its particular accretion history, related to the merging of massive
galaxies in each halo, and eventually the accretion of most of the
available gas and even other minor galaxies
\citep{cooray_dissipationless_2005}.
While galactic cannibalism has been proposed as the main mechanism for
building up central galaxies \citep{ostriker_another_1975,
white_dynamical_1976, vale_non-parametric_2006}, it has also been suggested
that major mergers \citep{lin_k-band_2004} and dry mergers
\citep{liu_major_2009,khochfar_dry_2009} play an important role in the
different stages of their evolution.
From a dynamic point of view, relative velocities of central galaxies with
respect to the halos in which they reside are very small, compared to the
large velocity dispersion of satellite galaxies.
This leads to a clear observational distinction between central and
satellite galaxies \citep{vale_non-parametric_2006}.
It is also known that a central galaxy is often the most luminous galaxy
within their halo \citep{vale_non-parametric_2006}.
Even when this ''central galaxy paradigm'' has been claimed to be
inaccurate \citep{skibba_brightest_2010}, specially in high mass systems,
the precise location of the center of the halo is not crucial for the
implementation of the method, since it integrates galaxy counts in a region
where overdensity signal is clearly present.
Still, the location of the brightest galaxy in this special type of galaxy
systems, where a galaxy strongly dominates in luminosity and the total mass
of the systems is quite low, is a very good approximation to the position
of the center of the galaxy system.
Under these assumptions, the brightest galaxies are expected to reside in
the centers of haloes \citep{jones_structure_1984,lin_k-band_2004,
smith_hubble_2005}, and can be used statistically to analyze galaxy
overdensities associated to satellites.
Accordingly, we limit the sample of centers to bright isolated galaxies,
since they are likely central galaxies of small haloes.

%--------------------------------------------------------- FIGURE 2 -
\begin{figure}
   \centering
   \includegraphics[width=0.49\textwidth]{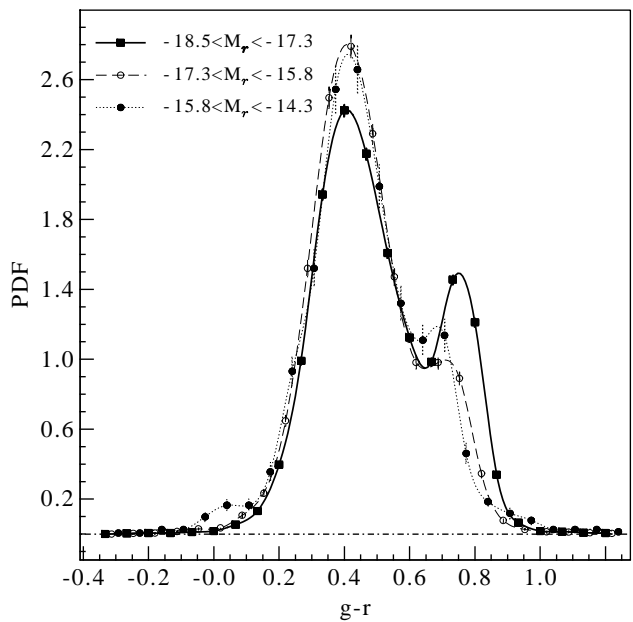}
   \caption{Colour distributions of faint galaxies with
   measured redshifts in SDSS, in three different absolute magnitude
   intervals.  Curves show spline interpolation.}
   \label{F_CD_spec} 
\end{figure}% 
%--------------------------------------------------------------------
 
%--------------------------------------------------------- FIGURE 3 -
\begin{figure}
  \centering
  \includegraphics[width=0.49\textwidth]{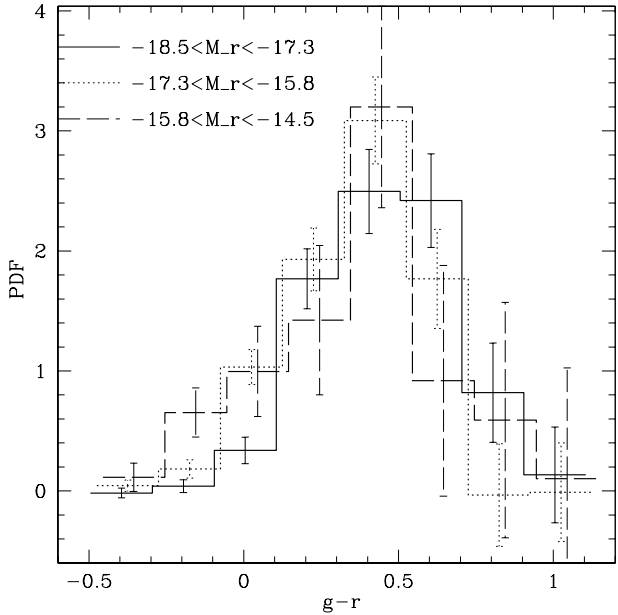}
   \caption{Colour distributions of excess counts of companions, in
  three different magnitude intervals, of bright primaries (sample
  S2-A) resulting from a background subtraction calculation.}
  \label{F_CRphoto} 
\end{figure}%
%--------------------------------------------------------------------

%}}}*/

\subsection{Constraints imposed by the data} \label{S_method:constraints}
%{{{SS*/

The main requirement for the success of a statistical background
subtraction method relies on a
significant mean overdensity around a given sample of primaries.
This fact leaded us to consider bright galaxies, \mbox{$M_p<-20.5$},
since the neighborhood of 
fainter primaries do not show a significant density enhancement
in SDSS data.
The uncertainty of the background decontamination procedure is
dominated by Poissonian statistics of subtraction of large numbers.
A significant increase in the signal to noise ratio of satellites
can be achieved by
eliminating those galaxies with a low probability of association to
the primaries.
To this end, we restrict the parameter space of galaxies in the
photometric catalogue so that we only use ranges of those parameters
where the hypothesis assumed in the background subtraction method are
best satisfied, and reliable estimations of luminosity, colour and
projected radial distributions can be obtained.
We impose constraints on apparent magnitude, colours and radial
distance of photometric galaxies.
Accordingly, we adopted $M_r<-14.$,
$-0.4<g-r<1.0$, and a projected radial distance in the range
$100$ kpc to $3<R_{vir}>$, which we discuss in detail in what follows.
 
The redshift range of primaries determines, along with the limiting
magnitude of the photometric sample, the maximum luminosity that can be
studied.
Given the limiting magnitude of $21.5$ in the r--band, and a minimum
redshift for the primaries of $0.03$, we chose the maximum luminosity 
$M_r-5log(h) \sim -14$ in the r--band.
The limiting magnitude we use is lesser than the limiting
magnitude from SDSS in order to ensure 100 per cent completeness
\citep{abazajian_second_2004}.

We show  in \fref{F_CD_spec} the 
observed colour distributions of SDSS satellite galaxies derived from the
spectroscopic sample in three different absolute magnitude bins between
$-18.5$ and $-14.5$.  It can be
appreciated in this figure 
that a colour cut in \mbox{$g-r<1$} includes most
of faint companion objects, and has the advantage of removing high redshift
galaxies reddened by K-correction.
Therefore, in all computations we exclude objects redder than this
threshold to lower the noise due to the presence of high redshift galaxies.
A clear indication that our colour cut \mbox{$g-r=1$} is a suitable
threshold to remove high redshift galaxies can be seen by inspection to
\fref{F_CRphoto} where we have applied background subtraction counts to
sample S2-A in the range \mbox{$100 $ kpc$<r_{p}<660$ kpc}.
The observed lack of excess signal beyond \mbox{$g-r=1$}  shows that our
method is effective in detecting companion galaxies in the colour range
\mbox{$-0.4<g-r<1$}.
This is also a convenient cut according to determinations of satellite
colours in semianalytic models of galaxy formation
\citep{font_colours_2008} and galaxies in groups in SDSS-DR2 data
\citep{weinmann_properties_2006}.
We have applied the same colour cut $g-r<1$ to galaxies
in the mock catalogue, as described in \sref{S_testing}.

\begin{figure}
   \centering
   \includegraphics[width=0.5\textwidth]{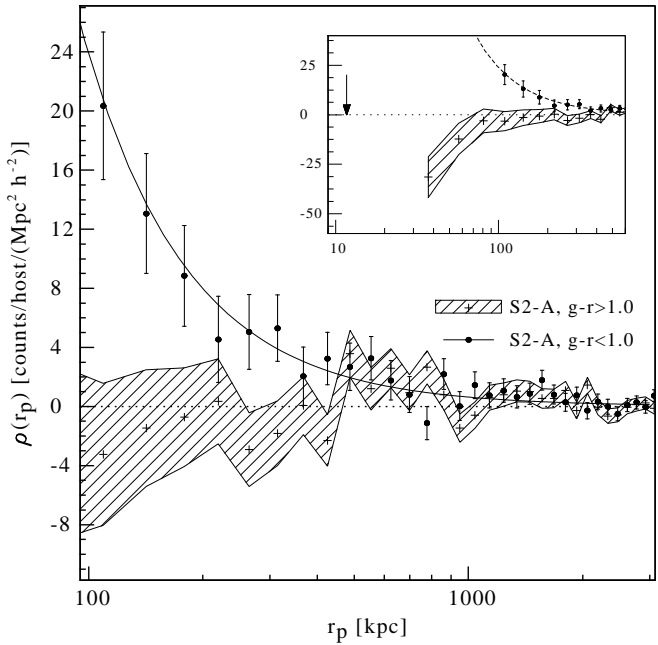}
 \caption{Mean projected density profile of galaxies with $g-r<1.0$
around primaries with $M_r<-21.5$ (sample S2).
Shaded region show the projected density profile of red galaxies
around the same primaries
with $g-r>1.0$ (mainly background objects), within 
1-$\sigma$ uncertainties.}
\label{F_DPandNoise}  
\end{figure}

In \fref{F_DPandNoise} we show the projected density profile of
galaxies, in the selected color range $-0.4 < g-r < 1.0$, around
primaries of sample S2-A-a.  For comparison, we also show the
corresponding of galaxies with $g-r>1$.
As can be appreciated, a significant excess of $-0.4 < g-r < 1.0$
galaxies is observed beyond $100$ kpc, while red galaxies, $g-r > 1.0$
(mainly consisting in strongly redshifted background galaxies) show a
null flat profile consistent with a uniform radial distribution
(shaded region in \fref{F_DPandNoise}), which also gives support to
our choice for the colour range of satellites.
However, as can be appreciated in the inset of \fref{F_DPandNoise}, we
notice that the projected density profile of red galaxies is not
uniform in the inmost region around the luminous primary galaxies.
This shows that the hypothesis of a uniform background fails within
$100$ kpc for this data set.  
Hence, in the computation of the statistical distributions, namely
luminosity and colours, galaxies are restricted to be at a projected
radial distance of at least $100$ kpc.
This radial distance is consistent with previous findings that
indicate the presence of 
an extended stellar halo associated to luminous galaxies.
This issue could lead to a failure of SDSS automated pipeline in the
detection of low surface brightness galaxies, either satellites or
foreground/background galaxies.
\citet{nierenberg_luminous_2011} study the spatial distribution of
faint satellites at intermediate redshifts ($0.1<z<0.8$) using high
resolution HST images of early type galaxies taken from GOODS fields.
The authors model the light profile of host galaxies in order to study
the population of faint satellites.  This model gives the spatial
distribution of satellites near the primaries as a combination of a
satellite population with a power law radial profile superimposed to
an isotropic and homogeneous background population.
The method proposed by the authors requires the analysis of images of
all hosts, which is beyond the scope of this work. 
This stellar halo component extends up to approximately $100$ kpc,
according to statistical determinations in SDSS
\citep{zibetti_halos_2003,bergvall_red_2009,tal_faint_2011}.

%--------------------------------------------------------- FIGURE 4 -
%-------------------------------------------------------------------- 

%--------------------------------------------------------- FIGURE 5
\begin{figure}
   \centering
   \includegraphics[width=0.49\textwidth]{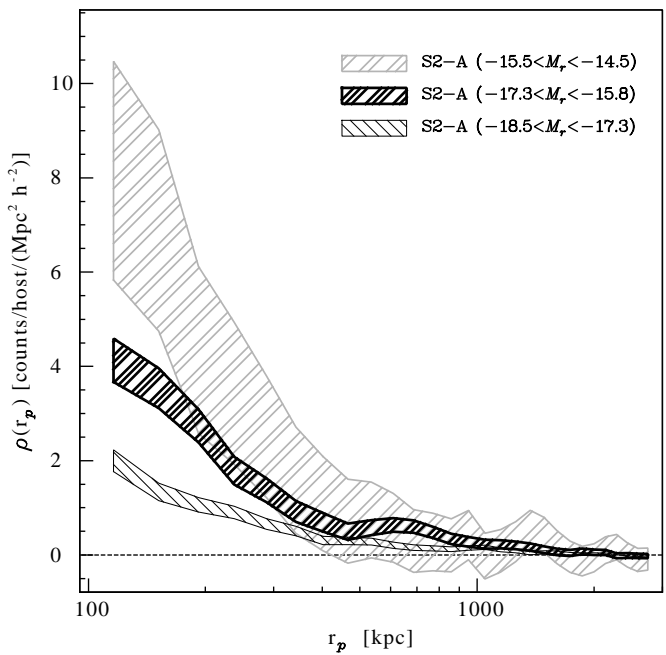}
   \caption{Density profile of satellites in S2-A sample selected 
   by satellite luminosity.  Galaxies are selected in the colour range 
   $-0.4<g-r<1.0$.
   Fainter satellites show a steeper and more
   concentrated density profile.
   Brighter satellites show a greater spatial extension, but smaller excess.}
   \label{F_DPbyMag}  
\end{figure}% 
%--------------------------------------------------------------------

An additional problem is related to fluctuations of luminosity density
the extended images of luminous galaxies, for example in the external
parts of large, late-type galaxies at low redshift.
Since SDSS images are analyzed to identify galaxies by automatic
algorithms, luminosity concentrations (which are part of the primary)
are often confused and classified as many faint fake galaxies.
This problem is hard to address, and is part of proposed SDSS
''research challenges"
\footnote{http://cas.sdss.org/dr7/en/proj/challenges/hii/default.asp}. 
The deblending process resulting from the current SDSS-DR7 pipeline
appears to induce an artificial neighbor excess, which strongly
difficults the treatment of data on regions close to the primary (see
e.g.  \citet{tollerund_smallscale_2011}), specially in late-type
galaxies where rich luminosity patterns are present on the disk.
We also show in the inset of \fref{F_DPandNoise} the mean value of the
$R_{90}$ distribution of primaries in sample S2, indicated with an
arrow.
Also, \citet{brainerd_anisotropic_2005} argue that satellites show a
strong preference for being aligned with the host major axis on scales
below $100$ kpc, while on larger scales ($250 \lesssim r_p/\rm{kpc}
\lesssim 500$) the satellite distribution is consistent with an
isotropic distribution.
Due to this anisotropy, satellites close and near the major axis of
the primary could be confused to luminosity enhancements of the disk
of the primary when using photometric data.
%
%The false identification of galaxies might be stronger in the
%proximity of the Petrosian $R_{90}$ radii of primaries, where the
%density profile artificially rises, while the external parts of the
%primary are affected by the stellar halo luminosity which blanket
%background galaxies.  
%
Both fake galaxies and depleted background are likely to coexist, and
can not be directly disentangled.
This might be also complicated by other effects such as dust
obscuration by primary disk, alignment of satellites with disks and
segregation of satellite properties with radial distance
\citep{chen_color_2007,ann_galactic_2008}.

The chosen minimum distance of \mbox{$100$ kpc} corresponds roughly to
\mbox{$0.5 - 0.6 R_{vir}$}.
\citet{tollerund_smallscale_2011} find that 12 per cent of Milky Way-like
galaxies host an LMC-like satellite within $75$ kpc projected
distance, while 42 per cent lie within $250$ kpc.
In the sample with the lower mean virial radius (S1) this means that,
for a typical primary galaxy, and an assumed power law profile of the
satellite distribution outside $20$ kpc ($\gamma=2.$, see
\tref{T_results}), only about 20 per cent of satellites within $100$ kpc are
not included in the analysis of sample S0 and $\sim 14$ per cent for sample
S2.

The projected density profile shows an excess of galaxy counts per
unit area, with respect to the local background, which decreases
beyond \mbox{$R_p \approx 500$ kpc}, but depends on the sample of
primaries and satellites considered.
\citet{sales_satellite_2005} analize simulations and find that the
turn around radius of satellite galaxies is of order $3\,R_{vir}$.
These findings are also consistent with the approximate location of
the turnaround radius according to the simple secondary infall model
\citet{bertschinger_self_1985}.
Then, in the computation of the statistical distributions, galaxies
are also restricted to be at a projected radial distance from the
primary of up to 3 times the mean virial radius in each sample. 
Using the masses of dark matter haloes hosting primary galaxies in a
mock catalogue (described in \sref{S_testing}), we find a variation in
the distributions of virial radius of primary galaxies corresponding
to samples S0, S1 and S2.
While the median virial radius $<R_{vir}>$ for sample S0 is
\mbox{$160$ kpc}, it changes to \mbox{$157$ kpc} for sample S1 and to
\mbox{$220$ kpc} for sample S2.
We adopt a maximum radius for each sample (S0, S1 and S2)
$R_{max}=3\,<R_{vir}>$, i.e., \mbox{$R_{max}=480$ kpc} for sample S0,
\mbox{$R_{max}=470$ kpc} for sample S1, and \mbox{$R_{max}=660$ kpc}
for sample S2. 
These values are also consistent with the results of
\citet{chen_constraining_2006} and \citet{liu_how_2010}, who find that
the fraction of interlopers remain low within $500 kpc$ projected
distance.
In \sref{S_testing} we use a mock catalogue to test the background
subtraction method, and confirm that the choice of this maximum radius
is convenient to maintain a low level of contamination by foreground
and background galaxies.

%--------------------------------------------------------- FIGURE 6 -
\begin{figure}
   \centering
   \includegraphics[width=0.49\textwidth]{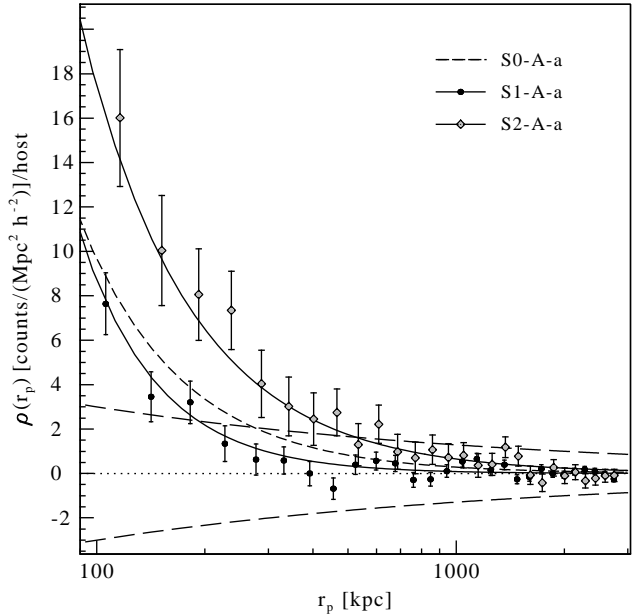}
   \caption{Density profile of excess galaxies around primaries in
   samples S0-A-a, S1-A-a, and S2-A-a.  For simplicity, only the power
   law fit is shown for sample S0-A-a, along with the uncertainty
   amplitude in long dashed lines.}
   \label{F_DP_ByLum}
\end{figure}%
%-------------------------------------------------------------------- 

In \fref{F_DPbyMag} we show the resulting profiles for different
ranges in satellite luminosities, where it can be seen that fainter
satellites are more strongly concentrated.
This maximum projected radius $3<R_{vir}>$ is suitable to study
satellite properties in the range $-18.5<M_r<-14.5$, although
brightest satellites can be detected beyond this value.
\citet{agustsson_orientation_2006} analyze simulations in a LCDM
framework to study the locations of satellite galaxies.
The authors define their samples of satellites as all galaxies around
luminous central galaxies, that have a difference in projected
distance lesser than 500kpc, and a difference in radial velocity below
500-1000 km/s, according to the selected sample.
Primary galaxies have a median host halo virial radius of ∼ 275 kpc.
\citet{ann_galactic_2008} analyze satellite galaxies in SDSS-DR5 in
order to study the dependence of radial distribution and environment
of galaxy morphology.  
These works show agreement on the extension of the satellite populations
up to
 $\sim 500$ kpc.

In order to account for
large scale angular fluctuations in the distribution of faint
galaxies we have used a local background in all samples.
The density profile is approximately constant beyond \mbox{$1.5$ Mpc}
(see \fref{F_DP_ByLum}), therefore we chose the density in the
projected radius interval \mbox{$2000-3000$ kpc} as the reference
level of background galaxy counts.
Although a global background could be used instead, we argue that a
mean number density of galaxies obtained locally is better suited to
account for possible irregularities in the number density of
background galaxies.
We use the redshift of each host to compute the absolute magnitudes
corresponding to the excess galaxy counts in different apparent
magnitudes, by assuming that these true companion galaxies are at the
same redshift than the primary.
We compute a composite luminosity distribution by averaging over a
sample of primaries with a given set of properties.
The number counts per magnitude bin on the resulting composite
luminosity distribution estimates are normalized according to the
appropriate limits of magnitudes in order to assure completeness:

\begin{equation}
  N_i = \sum_j{N_{ij} F_j C_j}
  \label{E_NumberSats}
\end{equation}

\noindent where $N_i$ is the number of galaxies within the $i$-th
magnitude bin for the total ensemble, and $N_{ij}$ is the number of
galaxies in that magnitude bin for the $j$-th primary.
The normalization for each magnitude bin is given by the fraction of
fields contributing within the completeness limits, $C_j$, and by the
fractional area, $F_j$, determined by the mask.

%---------------------------------------------------------- FIGURE 07 -
\begin{figure*}[!ht]
\centering
\hfill%
\subfigure[]{%
\label{F_mask}%
\includegraphics[width=.5\textwidth]{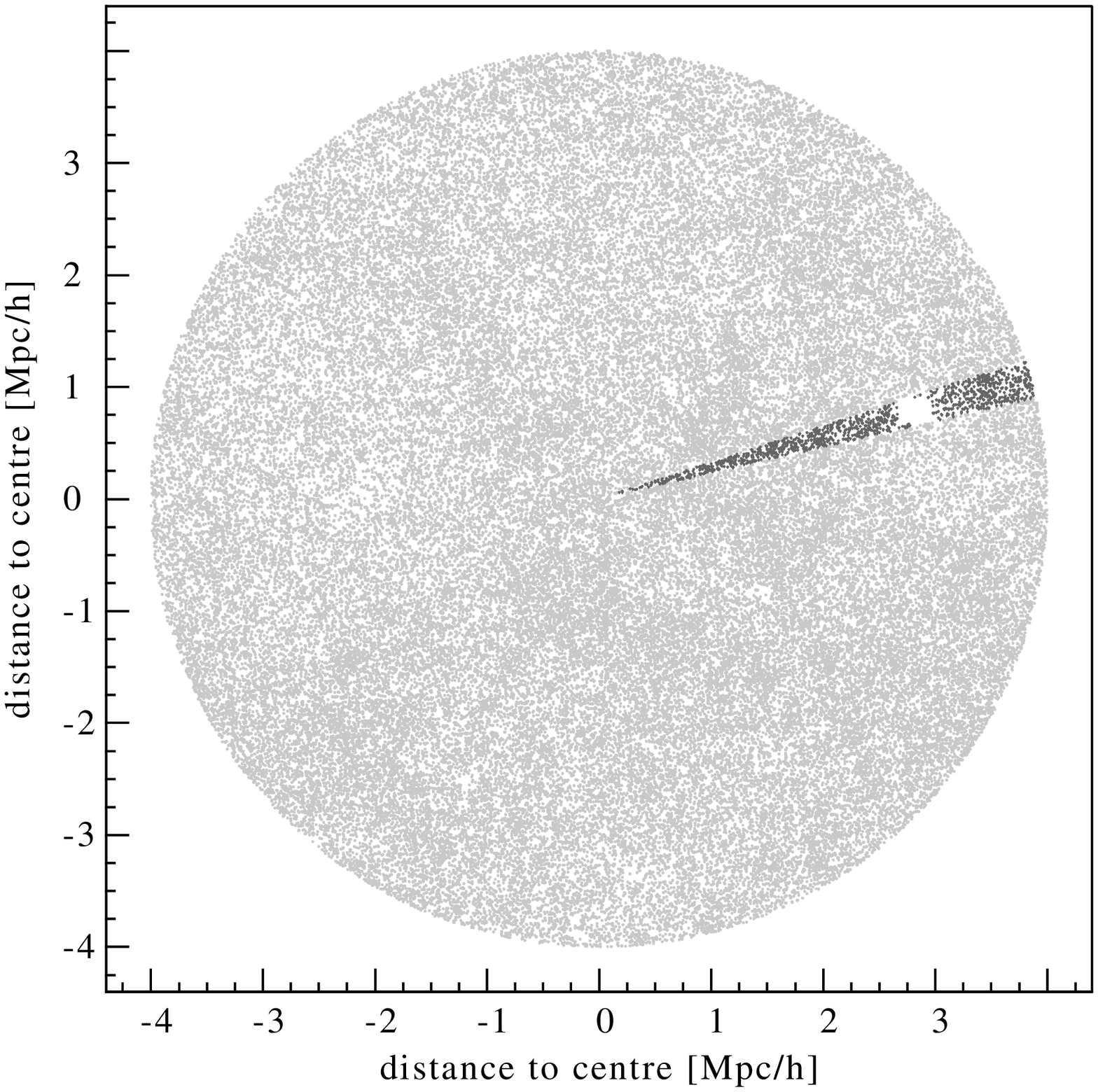}}~\hfill
\subfigure[]{%
\label{F_FracArea}%
\includegraphics[width=.5\textwidth]{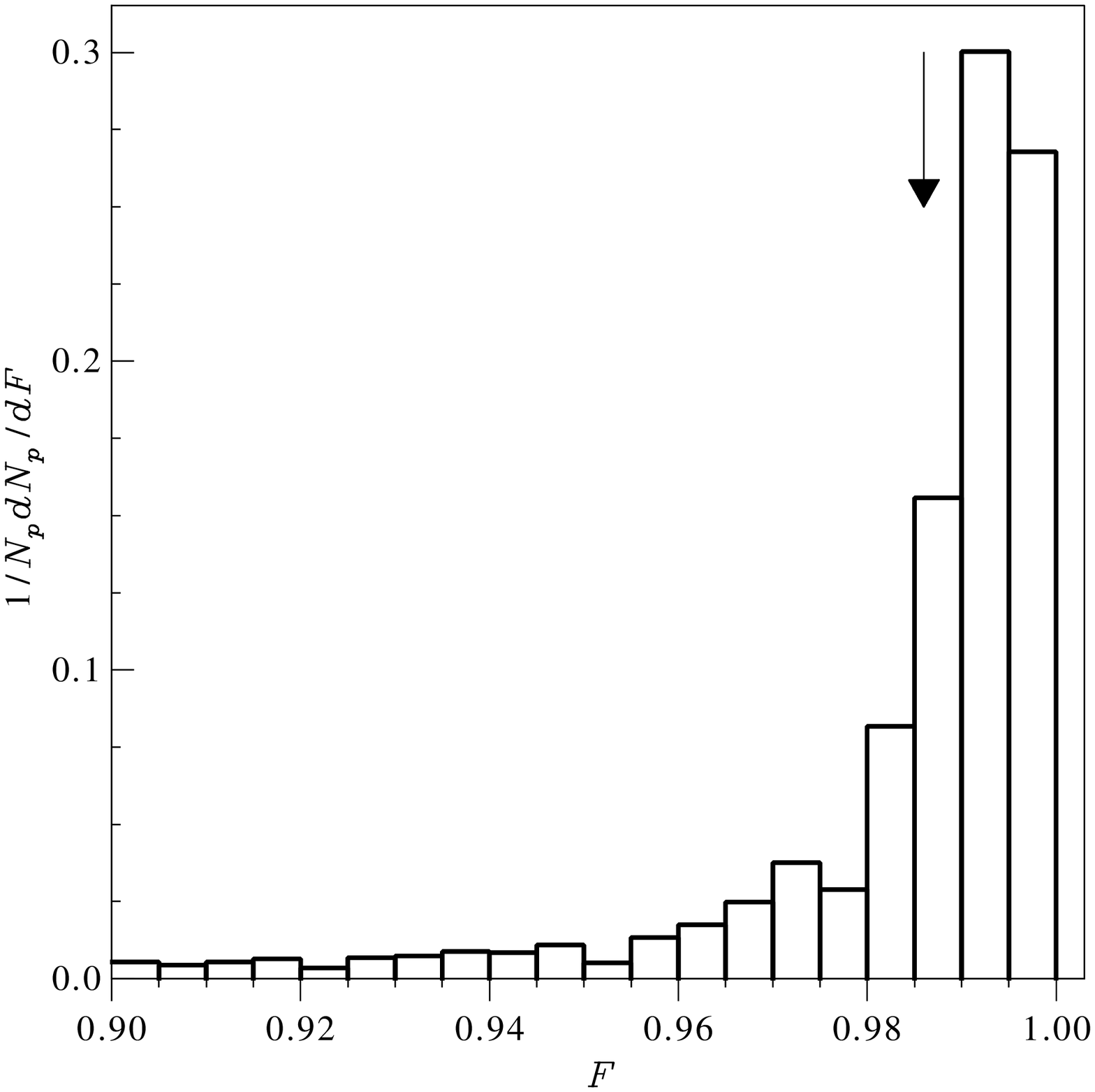}}\hfill~
\caption{%
   In left panel we show an example of a field centered 
   on a bright galaxy
   affected by a mask, where each point represents a
   galaxy in the photometric sample.
   Using the Monte Carlo procedure to construct a detailed mask, 
   as described in \sref{S_method:masks}, an angular fraction
   is eliminated (dark points) in the region where a hole is located.
   In this typical example, we use a 98.6 per cent of the area in
   the field.
   On the right panel, we show the 
   distribution of the fraction of areas of the primary fields
   $F$ after application of the masks.
   Fields with $F<0.9$ are excluded from the analysis.
   }
\end{figure*} 
%---------------------------------------------------------------------

%}}}SS*/

\subsection{Detailed masks} \label{S_method:masks}
%{{{/*

Since we expect a relatively small number of satellites compared to
the total background number counts, several primaries should be used
in order to add up the signal over the background noise.
Since the number of satellites grows approximately linearly with the
number of hosts and Poisson errors scale as the square root of the
number of galaxies, a large number of fields is required in order to
override the satellite population number over the background counts.
In \tref{T_SampleDefs} it can be seen that all samples comprise
thousands of primaries, and in particular, the smallest sample (S2-B)
contains nearly $5000$ host galaxies.
Fields of galaxies are extracted from the NYU-VAGC around the primary
galaxies.
The survey area from which the data is drawn has not, however, a
simple geometry, and the detailed structure of holes and borders is an
additional problem for the method used in this work. 
While the external borders are relatively simple, the survey has small
disconnected holes of a variety of shapes.
The origin of these holes relies on the presence of several bright
objects which blanket the background galaxies, in particular the faint
ones.
The most important are stars in our galaxy, but we can also mention
the trails of solar system objects or artificial satellites, cosmic
rays, etc.
Moreover, in certain cases, there is a regular pattern of holes,
sometimes associated to an incomplete structure of stripes.
Therefore, a mask must be built for each of the fields, in order to
correctly account for the effective areas used in the method.
As an example we show in \fref{F_mask} a typical field where
a hole in the projected distribution of galaxies can be seen.
The construction of the mask is based on the ansatz that, on a coarse
resolution, faint galaxies provide a dense, nearly uniform background
so that the holes can be directly associated to the absence of faint
objects in a patch of the sky.
In order to identify regions without faint galaxies we use a Monte
Carlo method.
The method consists in determining distances from the faint galaxies
in the photometric sample to a large number of uniformly distributed
random points within the area considered.
Those random points that are separated from its nearest neighbor
galaxy at least by a percolation radius $r_0$, are used to
characterize the holes, i.e., holes are detected as concentrations of
random points percolated by using this criteria.
We have adopted a percolation radius $r_0$ equal to a fraction $f$ of
the mean angular galaxy separation $<D>$ of each field.
The value of $f$ was obtained by diluting several realizations of
complete and dense fields of galaxies, and retrieving in each case the
ideal limiting value for the percolation radius.
We calibrated the value of this fraction,
and found a mean value of $f=0.17$, with a slight variation as a
function of the projected galaxy density in each field
\citep{lares_thesis}.
Given the complicated shapes of identified holes, we simply eliminate
angular sectors that contain the random points inside holes, as can be
appreciated in \fref{F_mask}.
In this Figure we show in grey points the locations of galaxies used
in the background subtraction procedure, and in dark grey points the
locations of galaxies in the same angular sector that the mask hole
that were not considered in such analysis.
We computed the distribution of the fractions of angular areas finally
used in the analysis presented in the following sections.
We show this distribution in \fref{F_FracArea}, where we have
indicated the location of the fraction of used area of the field shown
in \fref{F_mask}.
It can be seen from \fref{F_FracArea} the high level of completeness
of the data, with 90 per cent of the fields having less than 5 per
cent of their
areas lost by the application of the masks.
Furthermore, there is a negligible amount of fields with less than
90 per cent angular completeness, so that these fields are excluded from the
analysis.
Notice that this method is able to identify detailed small scale
features in the angular distribution of photometric galaxies.
Given the resolution required to account for small holes, a mask
constructed by using standard methods
\citep[e.g.][]{hamilton_mangle_2004, gorski_healpix_2005,
swanson_mangle2_2008} would be computationally expensive for reaching
the same accuracy than the used methodology.
Besides, this procedure allows to adapt the percolation radius for
each field, instead of using a fixed resolution on a pixelization
scheme, and makes the computation of areas straightforward by using
angular bins.

%}}}/*

\section{Results: statistical properties of satellite galaxies} 
\label{S_results}

 \subsection{Determination of projected radial distributions}
 \label{S_results:radial}
%{{{/*

Although it would be desirable to consider satellites as close as
possible to the primary galaxies, there are systematic detection
biases that strongly limit this possibility. 
We have also observed that close to primaries, the detection of faint
objects can be strongly biased due to several facts such as
obscuration, confusion, etc.
Galaxies behind bright galaxy discs could also be covered by intrinsic
absorption, producing significative changes in the observed
magnitudes.
Taking these issues into account, we have adopted a minimum distance
of \mbox{$100$ kpc} to primaries for our analysis assuring reliable
and systematic-free samples of faint objects.  
We considered galaxies in projected radial distance bins from the
corresponding primaries to compute the averaged density profile for
the different samples.
The radial bins are chosen so that all the resulting rings have the
same area.  This partition allows to explore the inner region with
more detail.
In \fref{F_DP_ByLum} we show the density profile of samples S1-A-a and
S2-A-a corresponding to all primaries with \mbox{$-21.5<M_r<-20.5$}
and \mbox{$-22.0<M_r<-21.5$} respectively, without restrictions in the
colours of primaries nor satellites.
It can be appreciated the smoothly declining mean density profiles,
although the extent of the overdense region is larger for the
brightest hosts.
We have also computed the density profiles for red
\mbox{($0.4<g-r<1.0$)} and blue \mbox{($g-r<0.4$)} companion galaxies
of bright primaries with \mbox{$M_r<-21.5$}, (samples \mbox{S2-A-r}
and \mbox{S2-A-b}).
It can be seen by inspection to \fref{F_DP_ByLum} that the two
profiles have a similar spatial extent and radial density profiles.

We have fitted power law functions to the radial density profiles
$\rho(r_p)=A r_p^{-\gamma}$.
The resulting values of the profile slope $\gamma$ are given in
\tref{T_results} where it can be seen that the companion galaxies of
primaries show in general a concentrated distribution.
\citet{agustsson_anisotropic_2010} investigate the locations of the
satellites of relatively isolated host galaxies in the Sloan Digital
Sky Survey and the Millennium Run simulation. 
The authors find that the distribution of the satellites within 500kpc
of red, high mass hosts with low star formation, differ from the
distribution of satellites of blue, low mass hosts with low star
formation.

%}}}/*

 \subsection{Mean number of satellites and luminosity distributions}
 \label{S_results:number}
 %{{{/*

%---------------------------------------------------------- FIGURE 08 -
\begin{figure}
    \centering                                            
    \includegraphics[width=0.45\textwidth]{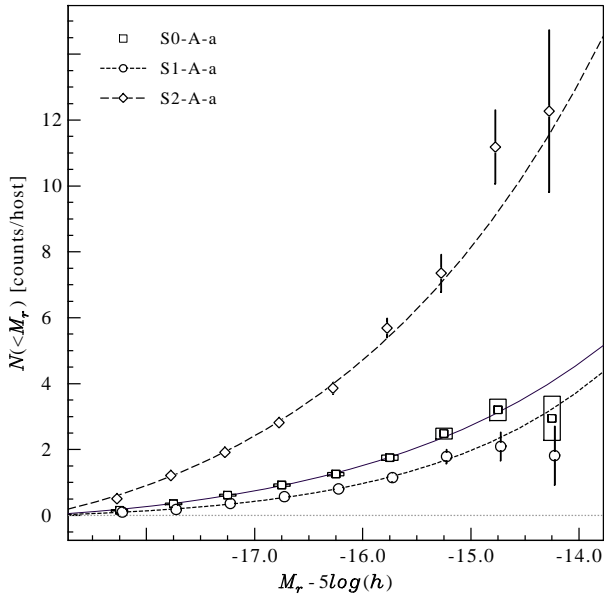}%
    \caption{
    Cumulative luminosity distributions of excess galaxies around
    primaries in the three defined samples (dubbed S0, S1 and S2).
    Lines represent the best Schechter function fits 
    to the obtained distributions (see text).
    }
    \label{F_LD_012AAAaaa}
\end{figure}
%---------------------------------------------------------------------

%---------------------------------------------------------- FIGURE 09 -
\begin{figure}
    \centering                                            
    \includegraphics[width=0.45\textwidth]{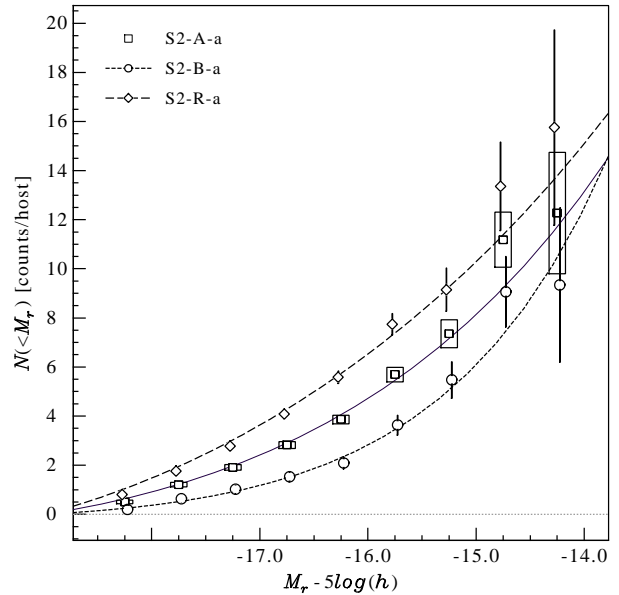}%
    \caption{
    Cumulative luminosity distributions of excess galaxies around
    bright primaries ($M_r<-21.5$).
    Luminosity distributions for red and blue hosts (with a colour cut
    in $g-r=1.0$) are also shown.
    }
    \label{F_LD_222ABRaaa}
\end{figure}    
%---------------------------------------------------------------------

We compute the mean number of satellites beyond $R_{min}=100$ kpc and
up to a projected radial distance equivalent to $R_{max}=3<R_{vir}>$
for each sample, no extremely red galaxies \mbox{($g-r>1$)} where
included, since both restrictions where applied in the computations as
previously explained (\sref{S_method:constraints}). 
The mean number of satellites (shown in \tref{T_results}) $<N_s>$ in
the magnitude range \mbox{$M_r<-14.5$} is estimated for all primary
samples, as:

\begin{equation}
  <N_s>=\frac{1}{N_p} \sum_{j=1}^{n} N_{in}^{j}-\frac{1}{A} N_{out}^{j},
\end{equation}  

\noindent where $N_p$ is the number of primaries in the sample,
$N_{in}^{j}$ is the number of galaxies inside the inner ring of the
$j$-th field, and $N_{out}^{j}$ is the number of galaxies inside the
chosen outer ring.  The area of the background, corrected by the mask,
is $A$ times the area enclosed between the chosen radius of the signal
inner region.
The isolation condition ensures that bright galaxies chosen as centers
are somewhat separated from other luminous galaxies, and is intended
for selecting a given group of galaxies with similar properties, but
does not affect the calculation of the luminosity function, given that
it includes satellites at most as bright as $M_s=M_p+2$
The faintest possible in sample S0 has an absolute magnitude
$M_p=-20.5$, so that no satellites with $M_s<-18.5$ are expected to be
found in the spectroscopic sample.
If galaxies brighter than this limit are considered, the sample of
satellites would not be complete, so we limit the study of satellites
to the range $-18.5<r<-14.5$.
We assign uncertainty estimates to the mean number of satellites using
Poisson errors resulting from the number counts of galaxies within
$R_{max}$ from the primaries and the expected number of background
galaxies:

\begin{equation}
\centering
   \epsilon = \frac{1}{N_p} \sqrt{\sum_{j=1}^{n}{N_{in}^{j}} + %
              \frac{1}{A} N_{out}^{j}}.
   \label{E_MeanNumberVariance}
\end{equation}

We find that the number of satellites depends on primary luminosity.
As can be seen in \tref{T_results}, sample S1 has on typically 1
satellite within the range of color and magnitude adopted.
This number increases to nearly 6 satellites on average for sample S2,
equally distributed between red and blue satellites.
This result is consistent with that of \citet{maccio_luminosity_2010},
who study numerical simulations of Milky Way sized haloes as predicted
by Cold Dark Matter based models of galaxy formation and find that the
number of satellite galaxies increase with halo mass.
On the other hand, the number of companion objects of primaries on the
same luminosity interval has not a clear dependence with satellite
$g-r$ colour index.
In all samples of primaries, the numbers of blue and red satellites are
comparable.
We notice that this is valid for the ranges of projected radius and
magnitudes used in this study (\tref{T_SampleDefs}).

In the computation of the luminosity distribution, a convenient
normalization is set for each magnitude bin, so that the mean number
of satellites per magnitude interval per primary is obtained.
This was achieved by dividing the number counts of excess galaxies by
the number of contributing primaries in each magnitude bin.
We performed Schechter fits \citep{schechter_analytic_1976} to the
differential histograms of magnitude distributions, and computed the
faint end slopes, given in \tref{T_results}.
For samples S0-A-a, S1-A-a and S2-A-a, the luminosity distributions
are consistent with a Schechter function with a faint--end slope of
$-1.3 \pm 0.2$, adopting a universal $M_{\star}$ value of $-20.44$
\citep{blanton_galaxy_2003}.

%------------------------------------------------------------ TABLE 2 -
\begin{table}
\begin{minipage}{\columnwidth}
\centering
\begin{tabular*}{\textwidth}{@{\extracolsep{\fill}}ccc}
%\begin{tabular}{cccc}

\hline
sample  &   $<N_{s}>$        &     $\gamma$ \\ 
\hline

S0-A-a  &  $1.5 \pm 0.4$ & -2.0$\pm$0.2  \\
S0-A-r  &  $0.7 \pm 0.4$ & -2.1$\pm$0.5  \\
S0-A-b  &  $0.9 \pm 0.2$ & -1.8$\pm$0.4  \\
\\                         
S1-A-a  &  $0.9 \pm 0.5$ & -2.4$\pm$0.3  \\
S1-A-r  &  $0.2 \pm 0.5$ & -2.0$\pm$2.0  \\
S1-A-b  &  $0.7 \pm 0.2$ & -1.9$\pm$0.5  \\
\\                         
S2-A-a  &  $6.1 \pm 1.4$ & -1.5$\pm$0.4  \\
S2-A-r  &  $3.4 \pm 1.3$ & -1.5$\pm$0.7  \\
S2-A-b  &  $2.7 \pm 0.6$ & -1.6$\pm$0.3  \\
\\                         
S2-R-a  &  $7.8 \pm 2.2$ & -1.1$\pm$0.6  \\
S2-R-r  &  $4.2 \pm 2.0$ & -0.8$\pm$0.7  \\
S2-R-b  &  $3.7 \pm 0.9$ & -1.8$\pm$0.2  \\
\\                         
S2-B-a  &  $4.7 \pm 1.8$ & -1.9$\pm$0.5  \\
S2-B-r  &  $2.5 \pm 1.6$ & -2.1$\pm$0.7  \\
S2-B-b  &  $2.1 \pm 1.0$ & -1.5$\pm$0.6  \\
\hline
\end{tabular*}
\end{minipage}
 
\caption{
   Mean number of satellites for each sample, and fit parameters for their
   background subtraction derived projected density profile.
   The mean number of satellites per host $<N_{s}>$ was
   calculated in the $r$--band. 
   Errors are calculated as described in \sref{S_method}.
   In all cases, primary redshifts lie in the range $0.03<z<0.10$ and
   the indicated number of satellites include objects in the magnitude
   range $-18.5<M<-14.5$, with the $g-r$ colour index limited to the
   range $-0.4<g-g<1.0$, and at a projected distance between $100kpc$
   and $R_{max}$ (see \tref{T_SampleDefs}).
}
\label{T_results} 
\end{table}
%--------------------------------------------------------------------

%--------------------------------------------------------- FIGURE 10 -
\begin{figure}
    \centering                                            
    \includegraphics[width=0.45\textwidth]{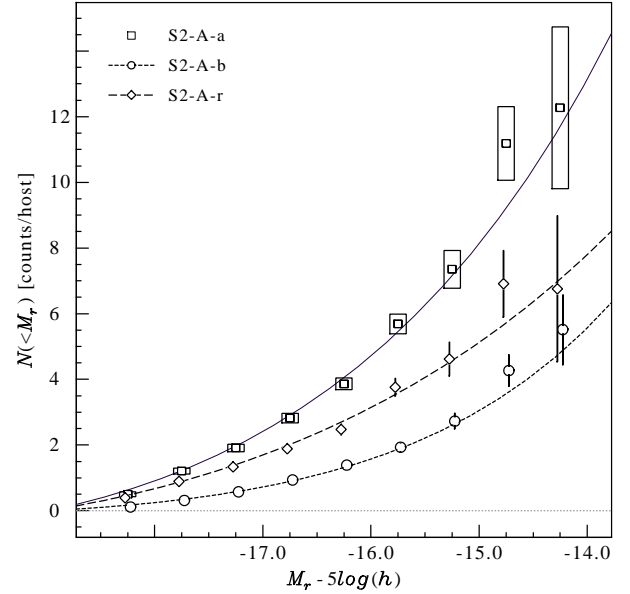}
    \caption{
    Cumulative luminosity distributions of 
    blue ($0.0<g-r<0.4$), red ($0.4<g-r<1.0$) and all ($0.0<g-r<1.0$)
    excess galaxies around bright primaries in sample S2-A.
    }
    \label{F_LD_222AAAabr}
\end{figure}
%---------------------------------------------------------------------

%---------------------------------------------------------- FIGURE 11 -
\begin{figure}
   \centering                                            
   \includegraphics[width=0.45\textwidth]{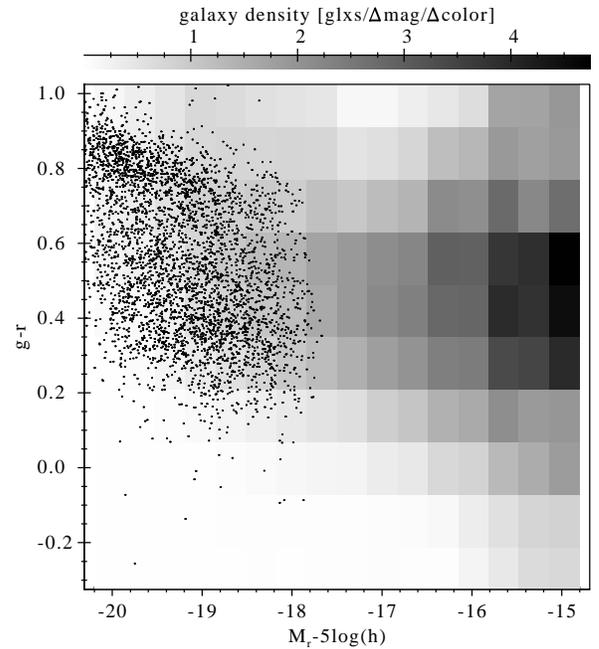}
   \caption{Colour-magnitude distribution of excess counts of 
   $-18<M_r<-15$ companions of bright primaries (sample S2-A)
   resulting from a background subtraction calculation.}
   \label{F_CM}
\end{figure}%
%-------------------------------------------------------------------- 

The results  shown in \tref{T_results} show a better signal to noise ratio
as more primaries comprise the subsamples.
Bright primaries \mbox{($M_r < -21.5$)} host on average 7 satellites while
low luminosity primaries \mbox{($-21.5 < M_r < -20.5$)} host a mean of only
1 satellite galaxy per primary. 
In \fref{F_LD_012AAAaaa} we show the $r$--band luminosity distribution of
satellites around bright primaries \mbox{($M_r<-21.5$)} and primaries with
intermediate luminosities (\mbox{$-21.5<M_r<-20.5$}, sample S1).
We also show in this Figure Schechter function fits computed using the
 universal value of $M_{\star}$, whith a faint end slope parameter $\alpha=-1.3 \pm 0.2$, obtained using a maximum likelihood method.
These luminosity functions indicate a lack of a dominant population of
faint satellites, which would be reflected in much larger negative values
of the $\alpha$ parameter.
These results contrast with those obtained for groups/clusters of galaxies
\citep{popesso_rass-sdss_2005, gonzalez_faint-end_2006}, where the faint
component contribution $M_r>-18.$ to a double Schechter fitting gives slope
values as steep as $\alpha\sim-2$.

We have also analyzed the dependence of the results on primary colour
index.
As mentioned in \sref{S_data:hosts} we have used the threshold
$g-r=0.8$ to divide the samples of primaries.
Similarly, we find a larger population of satellites associated to red
hosts, with a slightly stepper luminosity distribution at the faint end
(\fref{F_LD_222ABRaaa}).
We have studied the radial density profiles of red and blue satellites
around the different samples of primaries, finding that the system of blue
satellites is in all cases, more extended than that of the red ones by
approximately 30 per cent.
Luminosity distributions of red and blue satellites are shown for the
brightest hosts in \fref{F_LD_222AAAabr}.

We also computed a colour--magnitude diagram for the satellites obtained by
means of applying the background subtraction method simultaneously to these
two variables.
The results are shown in \fref{F_CM} for sample S2-A-a, where it can be
appreciated the smooth extension of the spectroscopic data onto fainter
objects obtained by our statistical approach. 
%}}}/*

\section{Testing the method with numerical simulations} \label{S_testing}
%{{{/*

The success of a background subtraction method relies on the
signal-to-background strength from satellites around bright galaxies,
providing the overdensity enhancement obtained in the stacking procedure.
However, the superposition of large scale structures projected onto the sky
could affect the uniformity of the background.
In order to estimate the ability of the method to correctly reproduce the
actual distributions of satellite galaxies in SDSS-DR7 data, we have tested
it on a mock catalogue derived from a numerical simulation using similar
conditions than that applied to the observations.
We constructed the mock catalogue within a $\pi/2$ sterradians light-cone,
based on a semianalytic model of galaxy formation \citep{croton_many_2006}
at redshift zero in the Millennium simulation
\citep{springel_simulations_2005}, which is publicly available from the
German Virtual Observatory \footnote{http://www.g-vo.org}.
This solid angle corresponds to approximately half the area covered by the
spectroscopic DR7 catalogue of galaxies and the $z=0$ snapshot was
replicated 8 times along the axes to achieve a suitable depth.

Since the output of the semianalytic model includes magnitudes in the
\textit{ugriz} photometric system, the mock spectroscopic catalogue is
obtained directly by selecting galaxies brighter than the limiting apparent
magnitude of the Sloan spectroscopic galaxy catalogue, \mbox{r=$17.77$}.
From the mock spectroscopic catalogue we extract a sample of mock primaries
using similar criteria than in \sref{S_data} which will be used as centers
in the following analysis.
For each galaxy, a redshift is assigned by placing a fiducial observer at
one corner, and determining the comoving distance to the observer and the
peculiar velocity of the galaxy.
The evolution corrected $r$--band magnitude is:

\begin{displaymath}
M_r = -2.5 \, \log (L) + E - 5 \, \log (h),
\end{displaymath}

\noindent
where $E$ is given by \citet{blanton_galaxy_2003}.
 
For the adopted r--band limiting magnitude $21.5$ of the photometric
sample, the maximum redshift of the mock should be $1.2$, corresponding
approximately to the distance at which an intrinsically luminous galaxy is
observable within the absolute magnitude range explored,
$-18.5<M_r<-14.5$.
Although the lack of evolution in both galaxies and structure is a drawback
of the mock catalogue, it serves as a strong test of the method given that
in this case there is a larger clustering amplitude of high redshift
structures (i.e. more structures along the line of sight) compared to a
mock catalogue with consistent evolution in clustering.

Since we have adopted a colour cut $g-r<1$ in SDSS data to reduce
background noise,  we have performed an appropriate noise reduction in the
mock catalogue using a Monte Carlo procedure in order to reproduce the
conditions of the observations.
We have considered photometric redshifts obtained by
\citet{omill_photoz_2010} (private communication) for the SDSS-DR7 galaxy
sample to derive the fraction $P(z)$ of galaxies with observed colour index
\mbox{$g-r<1$} as a function of redshift.
A suitable fit to this probability is given by $P(z)=1-2.381\,(z-0.08)$
which rejects about half the galaxies with $g-r >1$ at $z \sim 0.3$.
We have adopted this statistical procedure instead of a direct filtering of
galaxies by the observed colour in the photometric mock catalogue, given
that this would be model dependent, also requiring reliable K correction
for the semianalytic galaxies.
We also checked that our results are not strongly dependent on the precise
assumed $P(z)$, so that little modifications in the fit have minimum impact
in the obtained radial and luminosity distributions.

We have firstly tested if the galaxy density profile around primaries
derived by the background subtraction method is able to reproduce the
actual projected 3D profile.
For this aim we have computed the projected radial distribution of galaxies
around centers 
reproducing sample S2-A-a
in the mock catalogue using the real space positions in
order to test the reproducibility of the results through the background
subtraction method.
We show the results obtained for this sample (S2-A-a) in the mock catalogue 
since it presents the stronger signal.
The results described in this section, however, does not depend on the
chosen subsample.
The criteria to select satellites in the mock 3D catalogue, was chosen
so that a galaxy from the semi--analytic output is considered a satellite 
if it is within $3 R_{vir}$ of the host dark matter halo of the
primary galaxy, in three--dimensional space.
This is consistent with the criteria adopted in \sref{S_results:number}
for the selection of the region where the signal is studied in the 
observational samples.
In \fref{F_SimuVsMock_DP} we show the good agreement of the projected 3D
and background subtraction derived profiles, indicating that the method is
effective in recovering the true projected profile of companion galaxies.

The luminosity distribution of satellite galaxies were computed using the
halo membership and galaxy luminosities in the mock catalogue, and compared
to the luminosity distribution of galaxies obtained with the background
subtraction 
in the region $100<R_p/kpc<660$ corresponding to this sample.
The cumulative luminosity distribution obtained through the background
subtraction procedure reproduces the true underlying distribution
remarkably well, as can be appreciated in \fref{F_SimuVsMock_LD}.
We also display in the inset of this Figure the differential distributions,
which also present a general consistency.
Although a small difference is present between the two samples, 
that does not affect the shape of the luminosity distribution and the
determination of the Schechter parameter, 
since this is due to a small difference in the first magnitude bin.
The method also succeeds in reproducing the discontinuity at $M_r\approx
-17.6$, due to the absolute limiting magnitude in the parent semianalytic
galaxy catalogue.

These tests discard the possibility that background structures affect the
shape of the obtained luminosity distributions, so that the faint end
slopes are computed on firm statistical basis.
We stress the fact, however, that this procedure is reliable provided that
adequate selection criteria have been imposed to the data.
Since primary galaxies in our samples span a redshift range much smaller
than most of galaxies contributing to background counts, the procedure used
to compute radial and luminosity distributions performs in convenient
conditions, as our tests in the mock catalogue have shown.
This result is in conformity with previous tests of the method in less
favorable conditions as in the determination of cluster luminosity
functions \citep{valotto_clusters_2001,munoz_accuracy_2009}.

%}}}/*

\section{Discussion and conclusions} \label{S_conclusions}
%{{{/*

\begin{figure}
   \centering
   \includegraphics[width=0.49\textwidth]{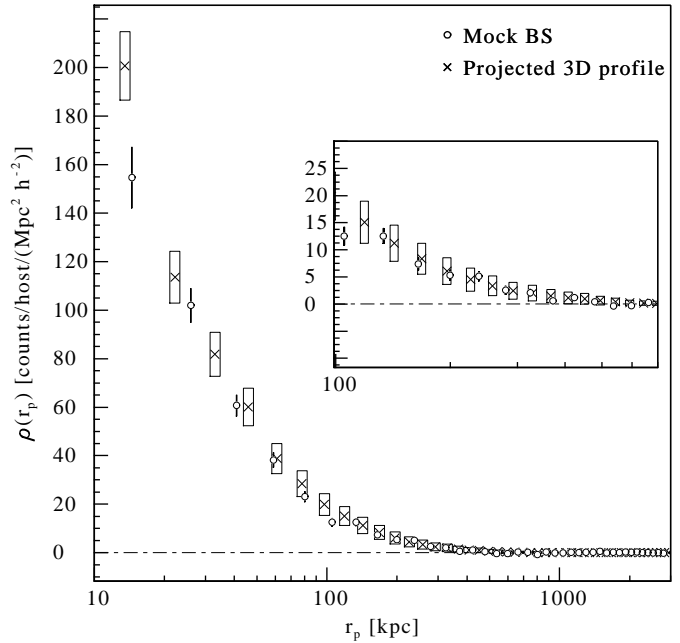}
   \caption{Projected density profile of galaxies around primaries
   in the mock catalogue.
   Open circles correspond to results obtained by the background
   subtraction method, and filled circles correspond to the projection
   of the 3D density profile within the halos of each primary.
   We show in the inset figure, the region we actually use in this
   work.
   Since the mock catalogue is not affected by spurious photometric 
   galaxy detections, we could also show the
   good agreement between both profiles at projected radius below
   \mbox{$100$ kpc}.
   }
   \label{F_SimuVsMock_DP}
\end{figure}

\begin{figure}
   \centering
   \includegraphics[width=0.5\textwidth]{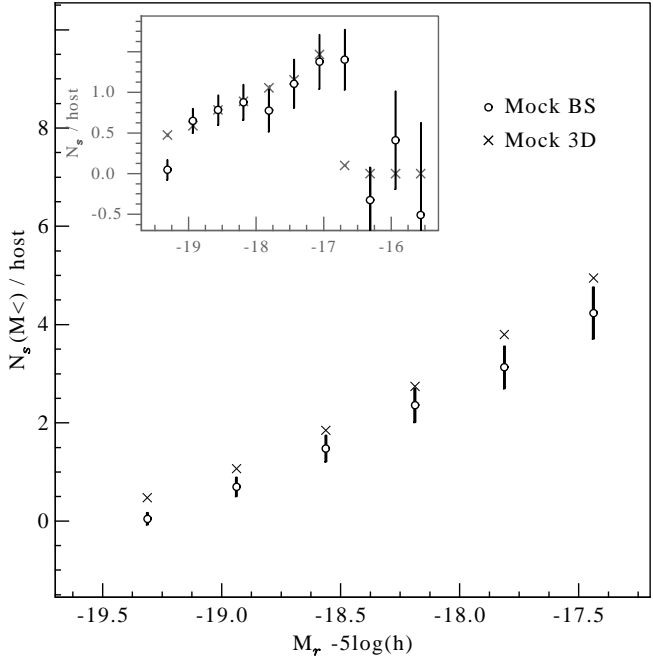}
   \caption{Cumulative luminosity distribution of 
   galaxies around primaries in the sample S2 of the mock catalogue.
   Open circles correspond to the results from the background
   subtraction method.  Filled circles are the actual luminosity
   distribution of satellites within three times the virial radius
   of the primary galaxies.
   In the inset we show the differential luminosity distribution
   on a luminosity range 
   beyond the completeness limit of the mock catalogue,
   where the background subtraction retrieve a distribution consistent
   with zero galaxies for $M_r>-17.$ in the Millennium semi--analytic
   catalogue.}
   \label{F_SimuVsMock_LD}
\end{figure}

We have carried out different statistical analyses to infer properties
of faint satellite galaxies 
in the projected distance range
\mbox{$100<r_p/\rm{kpc}<3<R_{vir}>$}, 
associated to bright primaries taken from SDSS with redshift $z<0.1$  
To this end, we have implemented a background subtraction method on
faint galaxies with photometric information, that are close in
projection to galaxies with measured redshifts.
The innermost region of the satellite systems is not
accessible using the proposed background subtraction method and data
from the photometric galaxy catalogue.  
However, assuming a power law profile for satellites and considering
objects outside 100 kpc radial distance from the host, we can study
more than 80 per cent of satellites.
We have used a mock galaxy catalogue based on a semianalytic model of
galaxy formation from the Millennium simulation to test the method with
similar conditions than in the observational data.  
According to the results of the tests performed, the method is able to
provide a good estimation of the true distributions of luminosities
and projected radial galaxy density as a function of the distances to
the host (\fref{F_SimuVsMock_DP} and \fref{F_SimuVsMock_LD}).
In our mock catalogue we test how the projection of background
structures affect our measurement in a worst case scenario.
We conclude that these structures does not affect the shape of the
luminosity distribution, provided that central galaxies are
sufficiently bright ($M_p<-20.5$) and isolated, and a colour cut
$g-r<1$ is imposed to satellites.
We also find it important to define an adequate maximum radius, using
theoretical insight and mock catalogues to calibrate the values of
$R_{max}$ for each sample.

%--------------------------------------------------------- FIGURE 12 -
%--------------------------------------------------------------------

%--------------------------------------------------------- FIGURE 13 -
%--------------------------------------------------------------------  

In all samples of primaries (defined in \tref{T_SampleDefs}) we detect an
excess of faint galaxy counts and we can determine statistically the
properties of companion objects associated to the central galaxy.
We find that the radial density profiles of satellites are consistent with
power laws of the form $\rho(r_p)=A r_p^{-\gamma}$, with \mbox{$-2.4
\lesssim\gamma\lesssim 1.1$} and that the maximum extent and amplitude of
the overdensity depends on the primary luminosity and colour (see
\fref{F_DP_ByLum}), as well as on galaxy luminosity (\fref{F_DPbyMag}).
The  dependence of the number of satellites with \mbox{$M_r<-14.5$} on host
luminosity is strong: bright primaries with $M_r<-21.5$ host on average
approximately 6 satellites, which is reduced to $\sim$1 satellite for
S1 primaries (\tref{T_results}).
\citet{liu_how_2010} investigate the probabilities of finding a Milky
Way like galaxy to host satellites with luminosities similar to the
Magellanic Clouds.
The authors report that $8$ per cent of these galaxies have at least 2
satellites similar to the LMC and SMC.
In order to compare with these results we computed the mean number of 
satellites in the magnitude range $-19.5<M_r<-14.5$ for 
Milky Way like primaries
 ($-21.4<M_r<-21.$), within $z=0.1$, using the previously described methodology.
We find a $(3\pm 2)$ per cent probability to obtain systems
similar to MW-LMC-SMC.
This is consistent with the results presented in
\citet{liu_how_2010}, since we exclude in the analysis 
the central \mbox{$100 kpc$} region.

Recently, \citet{wang_galaxy_2010} use a deep photometric
sample around spectroscopically identified galaxies, and found that
projected density profiles show a similar slope to the correlation
function slope, independently of galaxy luminosity.
Due to the different luminosity and redshift ranges considered between
the work of \citet{wang_galaxy_2010} and ours, a direct comparison is
difficult to perform.
While this paper was being reviewed, \citet{guo_satellite_2011}
presented a complementary analysis of the luminosities of satellites
of SDSS primary galaxies, finding a general agreement with our
results.

The redshift range of the spectroscopic sample and the apparent magnitude
limit of the photometric catalogue allows to obtain luminosity
distributions in the range $-18.5<M_r<-14.5$.
The derived luminosity distributions can be well described by Schechter
function fits (\fref{F_LD_012AAAaaa}).
Our findings indicate that faint end slopes of the satellite luminosity
functions are slightly rising ($\alpha=-1.3 \pm 0.2$).
This is in agreement with the luminosity distributions of galaxies in the
local group, in the same magnitude range \citep{mateo_dwarf_1998}.
This result is valid for all samples and indicates that the population of
satellites of bright isolated primaries are consistent with the nearly flat
faint end slope of the global luminosity function as derived for the SDSS
data \citep{blanton_galaxy_2003,baldry_sloan_2005,
montero-dorta_sdss_2009}.
These findings contrast with the results obtained by similar methods in
samples of clusters and groups where a significantly steep function is
obtained \citep[$\alpha \sim -2$ to $-1.5$, e.g.
][]{depropis_evidence_1995, popesso_rass-sdss_2005,
gonzalez_faint-end_2006}.
These results are expected, given the evidence from semianalytic models of
galaxy formation that suggest that the total mass of a dark matter halo
determines the normalization and shape of the luminosity function
\citep{maccio_luminosity_2010}.

%}}}/*

%··········································································

\acknowledgements 
%{{{/*
This work was partially supported by the Consejo Nacional de
Investigaciones Cient\'{\i}ficas y T\'ecnicas (CONICET), and the
Secretar\'{\i}a de Ciencia y Tecnolog\'{\i}a, Universidad Nacional de
C\'ordoba, Argentina.
We acknowledge Dr. Carlos Gutierrez and Dr. Nelson Padilla for useful
suggestions.
We acknowledge Lic. Ana O'Mill for providing data on photometric
redshifts.
We also thank the anonymous referee for
his/her through review and highly appreciate the comments and suggestions, 
which greatly improved this work.
We used the software HEALPix (http://healpix.jpl.nasa.gov/) 
to optimize the procesing of the galaxy catalogue.
We also used DISLIN plotting library and R statistical package.
Funding for the SDSS and SDSS-II has been provided by the Alfred P.
Sloan Foundation, the Participating Institutions, the National Science
Foundation, the U.S. Department of Energy, the National Aeronautics
and Space Administration, the Japanese Monbukagakusho, the Max Planck
Society, and the Higher Education Funding Council for England. The
SDSS Web Site is http://www.sdss.org/.
The SDSS is managed by the Astrophysical Research Consortium for the
Participating Institutions. The of the Royal Astronomical Society
Participating Institutions are the American Museum of Natural History,
Astrophysical Institute Potsdam, University of Basel, University of
Cambridge, Case Western Reserve University, University of Chicago,
Drexel University, Fermilab, the Institute for Advanced Study, the
Japan Participation Group, Johns Hopkins University, the Joint
Institute for Nuclear Astrophysics, the Kavli Institute for Particle
Astrophysics and Cosmology, the Korean Scientist Group, the Chinese
Academy of Sciences (LAMOST), Los Alamos National Laboratory, the
Max-Planck-Institute for Astronomy (MPIA), the Max-Planck-Institute
for Astrophysics (MPA), New Mexico State University, Ohio State
University, University of Pittsburgh, University of Portsmouth,
Princeton University, the United States Naval Observatory, and the
University of Washington.  
The Millenium Run simulation used in this paper was carried out by the
Virgo Supercomputing Consortium at the Computer Centre of the
Max--Planck Society in Garching.  The semi--analytic galaxy catalogue
is publicly available at
http://www.mpa-garching.mpg.de/galform/agnpaper.
% }}}/*

\label{lastpage}

\begin{thebibliography}{89}
\expandafter\ifx\csname natexlab\endcsname\relax\def\natexlab#1{#1}\fi

\bibitem[{{Abazajian} {et~al.}(2004){Abazajian}, {Adelman-McCarthy},
  {Ag{\"u}eros}, {Allam}, {Anderson}, {Anderson}, {Annis}, {Bahcall}, {Baldry},
  {Bastian}, {Berlind}, {Bernardi}, {Blanton}, {Bochanski}, {Boroski},
  {Briggs}, {Brinkmann}, {Brunner}, {Budav{\'a}ri}, \&
  {Carey}}]{abazajian_second_2004}
{Abazajian}, K., {et~al.} 2004, AJ, 128, 502

\bibitem[{{Abazajian} {et~al.}(2009){Abazajian}, {Adelman-McCarthy},
  {Ag{\"u}eros}, {Allam}, {Allende Prieto}, {An}, {Anderson}, {Anderson},
  {Annis}, {Bahcall}, \& et~al.}]{abazajian_seventh_2009}
{Abazajian}, K.~N., {et~al.} 2009, ApJS, 182, 543

\bibitem[{{Agustsson} \& {Brainerd}(2006)}]{agustsson_orientation_2006}
{Agustsson}, I., \& {Brainerd}, T.~G. 2006, ApJ, 644, L25

\bibitem[{{Agustsson} \& {Brainerd}(2010)}]{agustsson_anisotropic_2010}
---. 2010, ApJ, 709, 1321

\bibitem[{Andreon {et~al.}(2005)Andreon, Punzi, \&
  Grado}]{andreon_rigorous_2005}
Andreon, S., Punzi, G., \& Grado, A. 2005, MNRAS, 360, 727

\bibitem[{{Ann} {et~al.}(2008){Ann}, {Park}, \& {Choi}}]{ann_galactic_2008}
{Ann}, H.~B., {Park}, C., \& {Choi}, Y. 2008, MNRAS, 389, 86

\bibitem[{Baldry {et~al.}(2005)Baldry, Glazebrook, Budav\"ari, Eisenstein,
  Annis, Bahcall, Blanton, Brinkmann, Csabai, Heckman, Lin, Loveday, Nichol, \&
  Schneider}]{baldry_sloan_2005}
Baldry, I.~K., {et~al.} 2005, MNRAS, 358, 441

\bibitem[{Barkhouse {et~al.}(2007)Barkhouse, Yee, \&
  {López-Cruz}}]{barkhouse_luminosity_2007}
Barkhouse, W.~A., Yee, H. K.~C., \& {López-Cruz}, O. 2007, ApJ, 671, 1471

\bibitem[{Baugh(2006)}]{baugh_primer_2006}
Baugh, C.~M. 2006, Reports of Progress in Physics, 69, 3101

\bibitem[{{Benson}(2010)}]{benson_galaxy_2010}
{Benson}, A.~J. 2010, Phys. Rep., 495, 33

\bibitem[{{Benson} {et~al.}(2003){Benson}, {Bower}, {Frenk}, {Lacey}, {Baugh},
  \& {Cole}}]{benson_what_2003}
{Benson}, A.~J., {Bower}, R.~G., {Frenk}, C.~S., {Lacey}, C.~G., {Baugh},
  C.~M., \& {Cole}, S. 2003, ApJ, 599, 38

\bibitem[{{Bergvall} {et~al.}(2010){Bergvall}, {Zackrisson}, \&
  {Caldwell}}]{bergvall_red_2009}
{Bergvall}, N., {Zackrisson}, E., \& {Caldwell}, B. 2010, MNRAS, 405, 2697

\bibitem[{{Bertschinger}(1985)}]{bertschinger_self_1985}
{Bertschinger}, E. 1985, ApJS, 58, 39

\bibitem[{Bertschinger(1994)}]{bertschinger_cosmic_1994}
Bertschinger, E. 1994, Physica D Nonlinear Phenomena, 77, 354

\bibitem[{Blanton {et~al.}(2003{\natexlab{a}})Blanton, Lin, Lupton, Maley,
  Young, Zehavi, \& Loveday}]{blanton_efficient_2003}
Blanton, M.~R., Lin, H., Lupton, R.~H., Maley, F.~M., Young, N., Zehavi, I., \&
  Loveday, J. 2003{\natexlab{a}}, AJ, 125, 2276

\bibitem[{Blanton {et~al.}(2003{\natexlab{b}})Blanton, Hogg, Bahcall,
  Brinkmann, Britton, Connolly, Csabai, Fukugita, Loveday, Meiksin, Munn,
  Nichol, nori Okamura, Quinn, Schneider, Shimasaku, Strauss, Tegmark, Vogeley,
  \& Weinberg}]{blanton_galaxy_2003}
Blanton, M.~R., {et~al.} 2003{\natexlab{b}}, ApJ, 592, 819

\bibitem[{Blanton {et~al.}(2005)Blanton, Schlegel, Strauss, Brinkmann,
  Finkbeiner, Fukugita, Gunn, Hogg, Željko Ivezić, Knapp, Lupton, Munn,
  Schneider, Tegmark, \& Zehavi}]{blanton_new_2005}
---. 2005, AJ, 129, 2562

\bibitem[{{Brainerd}(2005)}]{brainerd_anisotropic_2005}
{Brainerd}, T.~G. 2005, ApJ, 628, L101

\bibitem[{{Chen}(2008)}]{chen_color_2007}
{Chen}, J. 2008, A\&A, 484, 347

\bibitem[{Chen {et~al.}(2006)Chen, Kravtsov, Prada, Sheldon, Klypin, Blanton,
  Brinkmann, \& Thakar}]{chen_constraining_2006}
Chen, J., Kravtsov, A.~V., Prada, F., Sheldon, E.~S., Klypin, A.~A., Blanton,
  M.~R., Brinkmann, J., \& Thakar, A.~R. 2006, ApJ, 647, 86

\bibitem[{Christlein(2000)}]{christlein_dependence_2000}
Christlein, D. 2000, ApJ, 545, 145

\bibitem[{{Coil} {et~al.}(2006){Coil}, {Gerke}, {Newman}, {Ma}, {Yan},
  {Cooper}, {Davis}, {Faber}, {Guhathakurta}, \& {Koo}}]{coil_deep2_2006}
{Coil}, A.~L., {et~al.} 2006, ApJ, 638, 668

\bibitem[{Cole {et~al.}(1994)Cole, {Aragon-Salamanca}, Frenk, Navarro, \&
  Zepf}]{cole_recipe_1994}
Cole, S., {Aragon-Salamanca}, A., Frenk, C.~S., Navarro, J.~F., \& Zepf, S.~E.
  1994, MNRAS, 271, 781

\bibitem[{{Collister} \& {Lahav}(2005)}]{collister_distribution_2005}
{Collister}, A.~A., \& {Lahav}, O. 2005, MNRAS, 361, 415

\bibitem[{Cooray \& Milosavljevic(2005)}]{cooray_dissipationless_2005}
Cooray, A., \& Milosavljevic, M. 2005, ApJ, 627, L85

\bibitem[{Cooray \& Sheth(2002)}]{cooray_halo_2002}
Cooray, A., \& Sheth, R. 2002, Physics Reports, 372, 1

\bibitem[{Croton {et~al.}(2006)Croton, Springel, White, Lucia, Frenk, Gao,
  Jenkins, Kauffmann, Navarro, \& Yoshida}]{croton_many_2006}
Croton, D.~J., {et~al.} 2006, MNRAS, 365, 11

\bibitem[{{de Propris} {et~al.}(1995){de Propris}, {Pritchet}, {Harris}, \&
  {McClure}}]{depropis_evidence_1995}
{de Propris}, R., {Pritchet}, C.~J., {Harris}, W.~E., \& {McClure}, R.~D. 1995,
  ApJ, 450, 534

\bibitem[{{Font} {et~al.}(2008){Font}, {Bower}, {McCarthy}, {Benson}, {Frenk},
  {Helly}, {Lacey}, {Baugh}, \& {Cole}}]{font_colours_2008}
{Font}, A.~S., {et~al.} 2008, MNRAS, 389, 1619

\bibitem[{Fukugita {et~al.}(1996)Fukugita, Ichikawa, Gunn, Doi, Shimasaku, \&
  Schneider}]{fukugita_sloan_1996}
Fukugita, M., Ichikawa, T., Gunn, J.~E., Doi, M., Shimasaku, K., \& Schneider,
  D.~P. 1996, AJ, 111, 1748

\bibitem[{Gonz\'alez {et~al.}(2006)Gonz\'alez, Lares, Lambas, \&
  Valotto}]{gonzalez_faint-end_2006}
Gonz\'alez, R.~E., Lares, M., Lambas, D.~G., \& Valotto, C. 2006, A\&A, 445, 51

\bibitem[{{G{\'o}rski} {et~al.}(2005){G{\'o}rski}, {Hivon}, {Banday},
  {Wandelt}, {Hansen}, {Reinecke}, \& {Bartelmann}}]{gorski_healpix_2005}
{G{\'o}rski}, K.~M., {Hivon}, E., {Banday}, A.~J., {Wandelt}, B.~D., {Hansen},
  F.~K., {Reinecke}, M., \& {Bartelmann}, M. 2005, ApJ, 622, 759

\bibitem[{Gunn {et~al.}(2006)Gunn, Siegmund, Mannery, Owen, Hull, Leger, \&
  Carey}]{gunn_2.5_2006}
Gunn, J.~E., Siegmund, W.~A., Mannery, E.~J., Owen, R.~E., Hull, C.~L., Leger,
  R.~F., \& Carey, L.~N. 2006, AJ, 131, 2332

\bibitem[{{Guo} {et~al.}(2011){Guo}, {Cole}, {Eke}, \&
  {Frenk}}]{guo_satellite_2011}
{Guo}, Q., {Cole}, S., {Eke}, V., \& {Frenk}, C. 2011, ArXiv e-prints

\bibitem[{{Hamilton} \& {Tegmark}(2004)}]{hamilton_mangle_2004}
{Hamilton}, A.~J.~S., \& {Tegmark}, M. 2004, MNRAS, 349, 115

\bibitem[{{Hansen} {et~al.}(2005){Hansen}, {McKay}, {Wechsler}, {Annis},
  {Sheldon}, \& {Kimball}}]{hansen_measurement_2005}
{Hansen}, S.~M., {McKay}, T.~A., {Wechsler}, R.~H., {Annis}, J., {Sheldon},
  E.~S., \& {Kimball}, A. 2005, ApJ, 633, 122

\bibitem[{Jones \& Forman(1984)}]{jones_structure_1984}
Jones, C., \& Forman, W. 1984, ApJ, 276, 38

\bibitem[{Kang {et~al.}(2006)Kang, Jing, \& Silk}]{kang_massive_2006}
Kang, X., Jing, Y.~P., \& Silk, J. 2006, ApJ, 648, 820

\bibitem[{{Khochfar} \& {Silk}(2009)}]{khochfar_dry_2009}
{Khochfar}, S., \& {Silk}, J. 2009, MNRAS, 397, 506

\bibitem[{Klypin {et~al.}(1999)Klypin, Kravtsov, Valenzuela, \&
  Prada}]{klypin_where_1999}
Klypin, A., Kravtsov, A.~V., Valenzuela, O., \& Prada, F. 1999, ApJ, 522, 82

\bibitem[{{Koposov} {et~al.}(2008){Koposov}, {Belokurov}, {Evans}, {Hewett},
  {Irwin}, {Gilmore}, {Zucker}, {Rix}, {Fellhauer}, {Bell}, \&
  {Glushkova}}]{koposov_luminosity_2008}
{Koposov}, S., {et~al.} 2008, ApJ, 686, 279

\bibitem[{Kravtsov {et~al.}(2004)Kravtsov, Gnedin, \&
  Klypin}]{kravtsov_tumultuous_2004}
Kravtsov, A.~V., Gnedin, O.~Y., \& Klypin, A.~A. 2004, ApJ, 609, 482

\bibitem[{{Lares}(2009)}]{lares_thesis}
{Lares}, M. 2009, PhD thesis, FaMAF, UNC

\bibitem[{{Li} {et~al.}(2007){Li}, {Jing}, {Kauffmann}, {B{\"o}rner}, {Kang},
  \& {Wang}}]{li_luminosity_2007}
{Li}, C., {Jing}, Y.~P., {Kauffmann}, G., {B{\"o}rner}, G., {Kang}, X., \&
  {Wang}, L. 2007, MNRAS, 376, 984

\bibitem[{{Lin} {et~al.}(2004){Lin}, {Mohr}, \& {Stanford}}]{lin_kband_2004}
{Lin}, Y., {Mohr}, J.~J., \& {Stanford}, S.~A. 2004, ApJ, 610, 745

\bibitem[{Lin {et~al.}(2004)Lin, Mohr, \& Stanford}]{lin_k-band_2004}
Lin, Y., Mohr, J.~J., \& Stanford, S.~A. 2004, ApJ, 610, 745

\bibitem[{{Liu} {et~al.}(2009){Liu}, {Mao}, {Deng}, {Xia}, \&
  {Wen}}]{liu_major_2009}
{Liu}, F.~S., {Mao}, S., {Deng}, Z.~G., {Xia}, X.~Y., \& {Wen}, Z.~L. 2009,
  MNRAS, 396, 2003

\bibitem[{{Liu} {et~al.}(2010){Liu}, {Gerke}, {Wechsler}, {Behroozi}, \&
  {Busha}}]{liu_how_2010}
{Liu}, L., {Gerke}, B.~F., {Wechsler}, R.~H., {Behroozi}, P.~S., \& {Busha},
  M.~T. 2010, ArXiv e-prints

\bibitem[{{Lupton} {et~al.}(2001){Lupton}, {Gunn}, {Ivezi{\'c}}, {Knapp}, \&
  {Kent}}]{lupton_SDSS_2001}
{Lupton}, R., {Gunn}, J.~E., {Ivezi{\'c}}, Z., {Knapp}, G.~R., \& {Kent}, S.
  2001, in Astronomical Society of the Pacific Conference Series, Vol. 238,
  Astronomical Data Analysis Software and Systems X, ed. {F.~R.~Harnden Jr.,
  F.~A.~Primini, \& H.~E.~Payne}, 269--+

\bibitem[{{Macci{\`o}} {et~al.}(2010){Macci{\`o}}, {Kang}, {Fontanot},
  {Somerville}, {Koposov}, \& {Monaco}}]{maccio_luminosity_2010}
{Macci{\`o}}, A.~V., {Kang}, X., {Fontanot}, F., {Somerville}, R.~S.,
  {Koposov}, S., \& {Monaco}, P. 2010, MNRAS, 402, 1995

\bibitem[{Mateo(1998)}]{mateo_dwarf_1998}
Mateo, M.~L. 1998, ARA\&A, 36, 435

\bibitem[{{Montero-Dorta} \& {Prada}(2009)}]{montero-dorta_sdss_2009}
{Montero-Dorta}, A.~D., \& {Prada}, F. 2009, MNRAS, 399, 1106

\bibitem[{Moore {et~al.}(1999)Moore, Ghigna, Governato, Lake, Quinn, Stadel, \&
  Tozzi}]{moore_dark_1999}
Moore, B., Ghigna, S., Governato, F., Lake, G., Quinn, T., Stadel, J., \&
  Tozzi, P. 1999, ApJ, 524, L19

\bibitem[{{Mu{\~n}oz} {et~al.}(2009){Mu{\~n}oz}, {Padilla}, \&
  {Barrientos}}]{munoz_accuracy_2009}
{Mu{\~n}oz}, R.~P., {Padilla}, N.~D., \& {Barrientos}, L.~F. 2009, MNRAS, 392,
  655

\bibitem[{Nichol {et~al.}(2003)Nichol, Miller, \& Goto}]{nichol_interplay_2003}
Nichol, R.~C., Miller, C.~J., \& Goto, T. 2003, apss, 285, 157

\bibitem[{{Nierenberg} {et~al.}(2011){Nierenberg}, {Auger}, {Treu}, {Marshall},
  \& {Fassnacht}}]{nierenberg_luminous_2011}
{Nierenberg}, A.~M., {Auger}, M.~W., {Treu}, T., {Marshall}, P.~J., \&
  {Fassnacht}, C.~D. 2011, ApJ, 731, 44

\bibitem[{Oemler(1974)}]{oemler_systematic_1974}
Oemler, A. 1974, ApJ, 194, 1

\bibitem[{{Okamoto} {et~al.}(2010){Okamoto}, {Frenk}, {Jenkins}, \&
  {Theuns}}]{okamoto_properties_2010}
{Okamoto}, T., {Frenk}, C.~S., {Jenkins}, A., \& {Theuns}, T. 2010, MNRAS, 406,
  208

\bibitem[{{O'Mill} {et~al.}(2010){O'Mill}, Duplancic, Lambas, \&
  Sodr\'e}]{omill_photoz_2010}
{O'Mill}, A., Duplancic, F., Lambas, D.~G., \& Sodr\'e, L. 2010, MNRAS,
  submitted to, 00

\bibitem[{Ostriker \& Tremaine(1975)}]{ostriker_another_1975}
Ostriker, J.~P., \& Tremaine, S.~D. 1975, ApJ, 202, L113

\bibitem[{Pimbblet(2008)}]{pimbblet_are_2008}
Pimbblet, K.~A. 2008, Are dumbbell brightest cluster members signposts to
  galaxy cluster activity?, {http://adsabs.harvard.edu/abs/2008arXiv0808.2093P}

\bibitem[{{Popesso} {et~al.}(2005){Popesso}, {B{\"o}hringer}, \&
  {Voges}}]{popesso_rass-sdss_2005}
{Popesso}, P., {B{\"o}hringer}, H., \& {Voges}, W. 2005, in Multiwavelength
  Mapping of Galaxy Formation and Evolution, ed. {A.~Renzini \& R.~Bender},
  444--+

\bibitem[{Sales \& Lambas(2005)}]{sales_satellite_2005}
Sales, L., \& Lambas, D.~G. 2005, MNRAS, 356, 1045

\bibitem[{{Schechter}(1976)}]{schechter_analytic_1976}
{Schechter}, P. 1976, ApJ, 203, 297

\bibitem[{Simon \& Geha(2007)}]{simon_kinematics_2007}
Simon, J.~D., \& Geha, M. 2007, ApJ, 670, 313

\bibitem[{{Skibba} {et~al.}(2010){Skibba}, {van den Bosch}, {Yang}, {More},
  {Mo}, \& {Fontanot}}]{skibba_brightest_2010}
{Skibba}, R.~A., {van den Bosch}, F.~C., {Yang}, X., {More}, S., {Mo}, H., \&
  {Fontanot}, F. 2010, MNRAS, 1465

\bibitem[{Smith {et~al.}(2005)Smith, Kneib, Smail, Mazzotta, Ebeling, \&
  Czoske}]{smith_hubble_2005}
Smith, G.~P., Kneib, J., Smail, I., Mazzotta, P., Ebeling, H., \& Czoske, O.
  2005, MNRAS, 359, 417

\bibitem[{Smith {et~al.}(2002)Smith, Tucker, Kent, Richmond, Fukugita,
  Ichikawa, ichi Ichikawa, Jorgensen, Uomoto, Gunn, Hamabe, Watanabe, Tolea,
  Henden, Annis, Pier, {McKay}, Brinkmann, Chen, Holtzman, Shimasaku, \&
  York}]{smith_ugriz_2002}
Smith, J.~A., {et~al.} 2002, AJ, 123, 2121

\bibitem[{Springel {et~al.}(2005)Springel, White, Jenkins, Frenk, Yoshida, Gao,
  Navarro, Thacker, Croton, Helly, Peacock, Cole, Thomas, Couchman, Evrard,
  Colberg, \& Pearce}]{springel_simulations_2005}
Springel, V., {et~al.} 2005, Nature, 435, 629

\bibitem[{Stoughton {et~al.}(2002)Stoughton, Lupton, Bernardi, Blanton, Burles,
  Castander, Connolly, \& Eisenstein}]{stoughton_sloan_2002}
Stoughton, C., Lupton, R.~H., Bernardi, M., Blanton, M.~R., Burles, S.,
  Castander, F.~J., Connolly, A.~J., \& Eisenstein, D.~J. 2002, AJ, 123, 485

\bibitem[{Strauss {et~al.}(2002)Strauss, Weinberg, Lupton, Narayanan, Annis,
  Bernardi, Blanton, \& Burles}]{strauss_spectroscopic_2002}
Strauss, M.~A., Weinberg, D.~H., Lupton, R.~H., Narayanan, V.~K., Annis, J.,
  Bernardi, M., Blanton, M., \& Burles, S. 2002, AJ, 124, 1810

\bibitem[{Strigari {et~al.}(2007)Strigari, Bullock, Kaplinghat, Diemand,
  Kuhlen, \& Madau}]{strigari_redefining_2007}
Strigari, L.~E., Bullock, J.~S., Kaplinghat, M., Diemand, J., Kuhlen, M., \&
  Madau, P. 2007, ApJ, 669, 676

\bibitem[{{Swanson} {et~al.}(2008){Swanson}, {Tegmark}, {Hamilton}, \&
  {Hill}}]{swanson_mangle2_2008}
{Swanson}, M.~E.~C., {Tegmark}, M., {Hamilton}, A.~J.~S., \& {Hill}, J.~C.
  2008, MNRAS, 387, 1391

\bibitem[{{Tal} \& {van Dokkum}(2011)}]{tal_faint_2011}
{Tal}, T., \& {van Dokkum}, P. 2011, 1102.4330

\bibitem[{{Tollerud} {et~al.}(2011){Tollerud}, {Boylan-Kolchin}, {Barton},
  {Bullock}, \& {Trinh}}]{tollerund_smallscale_2011}
{Tollerud}, E.~J., {Boylan-Kolchin}, M., {Barton}, E.~J., {Bullock}, J.~S., \&
  {Trinh}, C.~Q. 2011, ArXiv e-prints

\bibitem[{{Tollerud} {et~al.}(2008){Tollerud}, {Bullock}, {Strigari}, \&
  {Willman}}]{tollerund_hundreds_2008}
{Tollerud}, E.~J., {Bullock}, J.~S., {Strigari}, L.~E., \& {Willman}, B. 2008,
  ApJ, 688, 277

\bibitem[{{Trentham} \& {Tully}(2002)}]{trentham_faint_2002}
{Trentham}, N., \& {Tully}, R.~B. 2002, MNRAS, 335, 712

\bibitem[{{Trentham} \& {Tully}(2009)}]{trentham_dwraf_2009}
---. 2009, MNRAS, 398, 722

\bibitem[{{Tully} \& {Trentham}(2008)}]{tully_midlife_2008}
{Tully}, R.~B., \& {Trentham}, N. 2008, AJ, 135, 1488

\bibitem[{Vale \& Ostriker(2006)}]{vale_non-parametric_2006}
Vale, A., \& Ostriker, J.~P. 2006, MNRAS, 371, 1173

\bibitem[{Valotto {et~al.}(2001)Valotto, Moore, \&
  Lambas}]{valotto_clusters_2001}
Valotto, C.~A., Moore, B., \& Lambas, D.~G. 2001, ApJ, 546, 157

\bibitem[{Viola {et~al.}(2008)Viola, Monaco, Borgani, Murante, \&
  Tornatore}]{viola_how_2008}
Viola, M., Monaco, P., Borgani, S., Murante, G., \& Tornatore, L. 2008, MNRAS,
  383, 777

\bibitem[{{Wang} {et~al.}(2010){Wang}, {Jing}, {Li}, {Okumura}, \&
  {Han}}]{wang_galaxy_2010}
{Wang}, W., {Jing}, Y.~P., {Li}, C., {Okumura}, T., \& {Han}, J. 2010, ArXiv
  e-prints

\bibitem[{{Weinmann} {et~al.}(2006){Weinmann}, {van den Bosch}, {Yang}, \&
  {Mo}}]{weinmann_properties_2006}
{Weinmann}, S.~M., {van den Bosch}, F.~C., {Yang}, X., \& {Mo}, H.~J. 2006,
  MNRAS, 366, 2

\bibitem[{White(1976)}]{white_dynamical_1976}
White, S. D.~M. 1976, MNRAS, 174, 19

\bibitem[{White \& Frenk(1991)}]{white_galaxy_1991}
White, S. D.~M., \& Frenk, C.~S. 1991, ApJ, 379, 52

\bibitem[{Willman {et~al.}(2002)Willman, Dalcanton, Ivezic, Jackson, Lupton,
  Brinkmann, Hennessy, \& Hindsley}]{willman_sdss_2002}
Willman, B., Dalcanton, J., Ivezic, Z., Jackson, T., Lupton, R., Brinkmann, J.,
  Hennessy, G., \& Hindsley, R. 2002, ApJ, 123, 848

\bibitem[{{Yang} {et~al.}(2005){Yang}, {Mo}, {van den Bosch}, {Weinmann}, {Li},
  \& {Jing}}]{yang_cross_2005}
{Yang}, X., {Mo}, H.~J., {van den Bosch}, F.~C., {Weinmann}, S.~M., {Li}, C.,
  \& {Jing}, Y.~P. 2005, MNRAS, 362, 711

\bibitem[{{Zibetti} {et~al.}(2004){Zibetti}, {White}, \&
  {Brinkmann}}]{zibetti_halos_2003}
{Zibetti}, S., {White}, S.~D.~M., \& {Brinkmann}, J. 2004, MNRAS, 347, 556,
  {Mon.Not.Roy.Astron.Soc.347:556-568,2004}

\end{thebibliography}
\end{document}